\documentclass[12pt,a4paper]{article}
\usepackage[]{epsfig}	      
\usepackage[T1]{fontenc}      
\usepackage[latin1]{inputenc} 
\usepackage[english]{babel}    
\usepackage[]{latexsym}       
\begin{document}
\thispagestyle{empty}
\title{Exact electron states in $1D$ (quasi-) periodic 
arrays of delta-potentials.}
\author{Peter Kramer and Tobias Kramer\\
Institut f\"ur Theoretische Physik der Universit\"at \\
T\"ubingen, Germany.}
\maketitle

\section{Introduction and Scope.}

In the solid state physics of crystals, an important part is  
played by the band structure of the electronic states. This 
band structure arises from the representation of the periodicity
in the electronic state space. Powerful computational 
methods were developed for the calculation of band structures,
among them the linear muffin-tin (LMTO) method \cite{SKR} in the 
atomic sphere approximation (ASA) \cite{AND}. In the physics of quasicrystals,
it is believed that the electronic system plays an important part
\cite{TA}. Here one is lacking the periodic symmetry. To still use
the powerful methods of band computations, one must replace
the quasicrystal by a periodic approximant. The question arises 
how such approximant computations approach a quasiperiodic limit.
In a recent calculation \cite{HAE} it is shown by a supercell 
analysis that an approximant computation may lead to artefacts 
in the electronic density of states (DOS). More detailed 
local properties of the electrons appear as derivatives of the 
electronic charge density in Mössbauer studies on
quasicrystals \cite{KQS}.

In the present work we wish to analyse and compare electronic states in
periodic and quasiperiodic potentials in a way free of approximations.
In this way we can hope to address the similarities and differences of 
these systems from first principles.

To a first and very important approximation, the  many-electron states in 
ordered crystalline or quasicrystalline solids 
are constructed from one-electron states, compare for example
Ashcroft and Mermin \cite{AS},
after appropriate antisymmetrization.
The one-electron Hamiltonian then contains periodic and quasiperiodic 
potentials respectively. For a general account of band theory 
we refer to Blount \cite{BLO}. Analytic properties and the 
approach due to Wannier \cite{WAN} are discussed by Kohn
\cite{KOHN}. For general
electronic structures beyond periodicity, a local view was advocated by  Heine \cite{HEI} 
who proposed to throw out $K$-space. 

In what follows we wish to elaborate and to compare electronic  
systems and their local structure in terms of the following concepts:
\vspace{0.2cm}

(i) {\bf Local view of $K$-space in periodic potentials}: 

A Hamiltonian with an infinite {\em periodic potential} has the 
discrete symmetry group $\Lambda$ of its lattice. The electron states 
are characterized by the irreducible representations of $\Lambda$.
These are labelled by the continuous set of Bloch vectors
$K$ from the Brillouin zone, that is, the first 
unit cell of the reciprocal lattice $\Lambda^R$.
The eigenstates for fixed energy $E(K)$ are the Bloch states
belonging to the bands.

A local formulation of $K$-space arises as follows:
Wigner and Seitz \cite{WIG} in 1934 introduced the {\em cellular 
method}: The one-electron Schrödinger equation with fixed energy $E$
is to be solved exclusively on the finite unit
cell of the lattice with  the {\em local boundary conditions} that the 
solution, after propagation  over a full primitive period 
say $a_j,\; j=1,2,3$, picks up a 
pure phase factor. This phase factor when written in the form
$f_j := \exp (iK_j a_j),\; j=1,2,3$ determines the Bloch labels.
Two solutions with $K \rightarrow -K$ are degenerate. The energy 
$E(K)$ appears in  bands, with Bloch labels ranging over the
Brillouin zone, and in gaps where there are no states with Bloch
type boundary conditions. This approach allows an extension to finite 
and to infinite systems. Within bands,
the solutions can be  matched and propagated as Bloch states 
over all the cells of the crystal.\vspace{0.2cm} 
 
(ii) {\bf Local view of $K$-space in quasiperiodic potentials}:

For infinite almost periodic systems
in $1D$ some general results with many references are discussed in  \cite{CY}.
For another  detailed study of discrete systems with many references we 
quote Sütö \cite{SU}.
For our purposes we consider an infinite {\em quasiperiodic potential} built
from a few basic cells as in a tiling model for quasicrystals.
For each cell one can apply the local analysis of Wigner and Seitz
with local Bloch type boundary conditions. Quasiperiodicity now requires
a matching of these local boundary conditions between the basic cells.
It is not enough that, at fixed energy $E$, each cell separately admits
Bloch type boundary conditions. We must also check if a patch of 
cells admits these boundary conditions. Again one can  study
the quasiperiodic extension of the system, following for 
example a geometric inflation of the tiling.\vspace{0.2cm}

(iii) {\bf Local view, positive and negative energy, motive clusters}:

Imagine the atomic nuclei of a unit cell in a periodic crystal 
to form a free {\em motive cluster}. This motive cluster is not unique 
since the unit cell of a crystal admits translations as well 
as transformation of 
shape, for example in going from the primitive to the Wigner-Seitz cell.

For a fixed choice of motive clusters we should not  relax the positions of the nuclei 
to gain energy,
as would be done in real  molecules, since in the crystal 
the positions are controlled
by {\em all} atomic neighbours.

{\em Example}: Take for example a linear crystal with two atoms $A,B$, 
in two spacings 
$a \neq b$ 
with the sequence AaBbAaBbAa... and period $a+b$. Choose 
two different motive clusters by the 
replacements $AaBb \rightarrow AaB$ and $BbAa \rightarrow BbA$ 
which do not affect the bound states. 
We  get in general two different bound states, with the binding energy
depending on the relative distances $a$ and $b$ respectively.
In a linear molecule we would relax the distance and end up in both cases
at the same binding energy. This is not allowed in the crystal!
In the linear crystal {\em the  two different motive clusters generate  
identical  periodic structures and bands}!

Consider then the one-electron bound states for the chosen motive cluster. 
It is true that, in order to treat it like a molecule, we should take along
the nuclei and their Coulomb repulsion. However since we keep the nuclei at fixed
positions we may neclect these interactions which of course influence the 
binding and the total energy. Independent of these interaction terms,
the different boundary conditions of the electron states at positive or negative
energy will enter into the computation of the many-electron states.

Since the Hamiltonian for fixed nuclear positions
is the same as in the Wigner-Seitz method, the only difference
is in the boundary conditions which for a bound state require 
the solutions to decay exponentially. Clearly now we must distinguish
between the {\em positive and negative energy region}. The former 
admits scattering states,  the latter 
bound states. 

The infinite systems at positive energy should here be considered as an 
approximation
to a finite macroscopic systems which still admits a scattering experiment.
We could also consider half-infinite systems which allow at least
for backscattering. For both finite and infinite systems, the distinction
between positive and negative energy states is crucial. In considering
finite patches of periodic or quasiperiodic systems one therefore should pay attention
to the sign of the energy.\vspace{0.2cm}

(iv){\bf Bound, Bloch and scattering states in periodic potentials}:

We shall term the negative energy   region the 
{\em exact tight binding scenario} in what follows.
As a partial differential equation of second
order, the stationary Schrödinger equation has two fundamental 
solutions on a single cell. Suppose that the energy of a bound
state lies within the range of energy for a band. Then it follows 
from the Schrödinger equation that
{\em the bound state solution restricted to the unit cell must be
an exact linear combination of the local Bloch states 
at the bound state energy}.
In the positive energy region we can still obtain states with 
local Bloch boundary conditions. Now we can inquire 
how the band structure of these positive energy states is related to one-electron scattering
states.

{\em Example}: The relation of bound and Bloch states in crystals is often 
considered in approximate rather than exact 
computations, as for example in the {\em 
tight binding approximation}. Here one starts exclusively 
from bound atomic l-orbitals 
of the isolated atoms, obtains the Bloch states from their 
tunneling to next neighbours, 
and labels the bands by the l-orbitals. If first-order
degenerate perturbation theory applies, the energies in the bands
result from the splitting of the  single orbital whose degeneracy is
proportional to the number of atoms.
In this approximation it is clear that the energy of 
the local atomic orbital is within the corresponding band.
In many cases the bound atomic orbitals of the atoms in the cell
are separated in energy. Otherwise one can consider hybrid or 
motive 
cluster orbitals on the unit cell as the origin of band labels.
The discrete version of this approximation leads to an eigenvalue
problem and therefore looses the information on the sign
of the energy.

The general arguments given above are valid  {\em independent
of any such approximation scheme}. They  show that {\em in the exact 
tight binding scenario there must exists an exact local 
relation between bound 
and Bloch states}.\vspace{0.2cm}

(v){\bf Bound, Bloch and scattering states in quasiperiodic potentials}:

For a quasiperiodic potential based on a tiling, we again wish to 
distinguish the negative and positive energy regions. One can 
consider at negative energy the bound states for finite 
patches which form part of a quasiperiodic structure
and compare  with states obeying local Bloch type boundary conditions 
on the same patch. We refer to such an energy range as a 
{\em band germ} and to the local Bloch states as {\em Bloch germs}. 
If the bound state energy admits Bloch type boundary conditions,
it must be possible to express the bound states by pairs of 
local Bloch states (germs). The bound states may be related to
local clusters.

In the scattering from quasiperiodic potentials at
positive energy, the aim is again to compute the scattering matrix.
For an exact computation one may inquire about the implication of
inflation symmetry on the scattering matrix.
\vspace{0.2cm}
  
In what follows we wish to demonstrate the validity of these concepts in a way 
free of approximations.
In dimension three we cannot yet implement exact examples for
these concepts. The LMTO method may allow to examine some of our
points.
We choose here   a comparative and specific 
study restricted to  
continuous periodic and quasiperiodic potentials in dimension one. 
On them we wish to demonstrate the concepts given above by exact
computations.
To do so we pay the price 
of choosing simple delta potentials even of the same strength.  
We explore the electron states in terms of local boundary conditions,
in the spirit of the Wigner-Seitz  approach, applied to finite 
strings (patches) which form part of periodic or quasiperiodic structures.
We shall always start from finite patches and then look into their 
recursive extension.
The solutions with Bloch  boundary conditions on a finite string are  
systematically compared with bound state solutions on the same string.

The main tool of our analysis is the continuous transfer matrix
which is well defined for piecewise constant potentials,
including delta potentials  as a limiting case.
This matrix propagates by matrix multiplication 
a fundamental system of two solutions. We shall stress in what follows
the polynomial dependence of the matrix elements on the strength
parameter of the delta potential. This will allow us to order
and analyze the matrix systems.

\section{Finite periodic strings at negative energy.}

\subsection{Preview: An energy gauge for crystals.}

As an introduction we present some material from \cite{KR3}.
We rephrase the well-known periodic case \cite{LIE}.
Consider first a finite string $S$ with
the transfer matrix $M$, described in more detail in section 2.2. 
We define an {\em energy gauge} $f$ as a function
with value $f=0$ on 
an energy interval such that $|tr(M)| \leq 1$
and $f=1$ otherwise, compare Fig. 1. We call this 
a {\em band germ}. The condition on the trace assures that
inside the band germ $M$ has two complex conjugate eigenvalues 
of absolute value $1$.
Repeat the string $n$ times to produce 
a new  transfer matrix $M^n$. Since the existence of band germs 
is related to the eigenvalue problem, the range of energies 
for which $\frac{1}{2}|tr(M^n)|<1$ is {\em independent of} $n$,  
{\em the band germs stay the same on the new string}. 
The Bloch germs propagate through the string and pick
up the same phase factors respectively 
after each transmission. 

Consider next the bound states of the string $M^n$.
In a very tight binding case  we claim 

${\bf 1 \, Prop}$: The bound states of the string $M^n$ 
may be grouped into sets of n
states, where the energies of each such set corresponds to
a part $f=0$ of the energy gauge $f$ of the string $M$. 

{\em Proof}: For very tight binding 
it suffices to use first-order degenerate perturbation theory, applied to the 
bound states of the single atoms: To this order,
the bound states of the string are the eigenvalues of 
a matrix whose off-diagonal entries are  the weak atom-atom cross terms.
By standard matrix theory, the maximum level splitting will increase 
with $n$. But band theory tells us that in  the limit 
$n \rightarrow \infty$ all energies stay inside the band, hence inside
the initial band germ. Thus 
the energy of all these bound states for finite $n$ must stay 
within the energy gauge
of the (single) band germ. 

We now have the following situation in the finite string $M^n$: 
In a band germ from the string $M$ , there is for each energy value a pair of 
Bloch germs
which can carry  charge current. There are now $n$ discrete bound states
with the energy gauge as in the initial string.

\begin{center}
\begin{picture}(0,0)%
\includegraphics{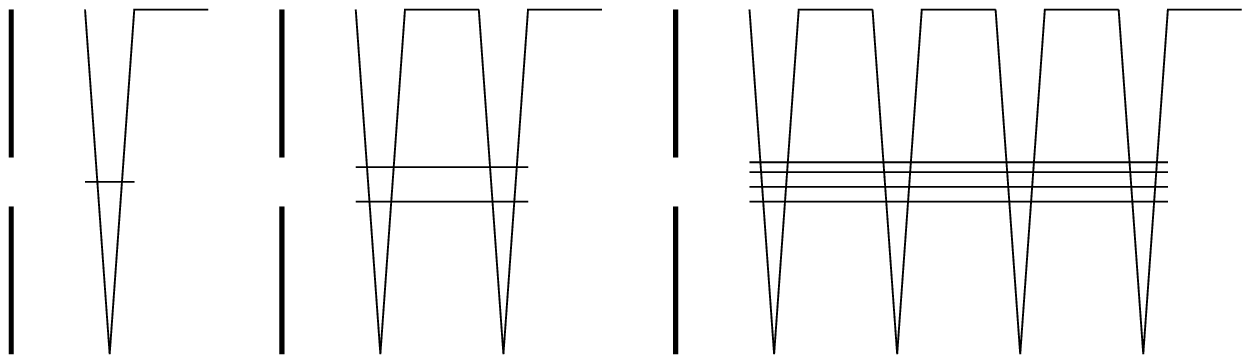}%
\end{picture}%
\setlength{\unitlength}{2072sp}%
\begingroup\makeatletter\ifx\SetFigFont\undefined
\def\x#1#2#3#4#5#6#7\relax{\def\x{#1#2#3#4#5#6}}%
\expandafter\x\fmtname xxxxxx\relax \def\y{splain}%
\ifx\x\y   
\gdef\SetFigFont#1#2#3{%
  \ifnum #1<17\tiny\else \ifnum #1<20\small\else
  \ifnum #1<24\normalsize\else \ifnum #1<29\large\else
  \ifnum #1<34\Large\else \ifnum #1<41\LARGE\else
     \huge\fi\fi\fi\fi\fi\fi
  \csname #3\endcsname}%
\else
\gdef\SetFigFont#1#2#3{\begingroup
  \count@#1\relax \ifnum 25<\count@\count@25\fi
  \def\x{\endgroup\@setsize\SetFigFont{#2pt}}%
  \expandafter\x
    \csname \romannumeral\the\count@ pt\expandafter\endcsname
    \csname @\romannumeral\the\count@ pt\endcsname
  \csname #3\endcsname}%
\fi
\fi\endgroup
\begin{picture}(11497,3800)(451,-3705)
\put(451,-61){\makebox(0,0)[lb]{\smash{\SetFigFont{12}{14.4}{rm}$f(S)$}}}
\put(1351,-61){\makebox(0,0)[lb]{\smash{\SetFigFont{12}{14.4}{rm}$S$}}}
\put(2701,-61){\makebox(0,0)[lb]{\smash{\SetFigFont{12}{14.4}{rm}$f(S^2)$}}}
\put(3826,-61){\makebox(0,0)[lb]{\smash{\SetFigFont{12}{14.4}{rm}$S^2$}}}
\put(6301,-61){\makebox(0,0)[lb]{\smash{\SetFigFont{12}{14.4}{rm}$f(S^4)$}}}
\put(7426,-61){\makebox(0,0)[lb]{\smash{\SetFigFont{12}{14.4}{rm}$S^4$}}}
\end{picture}

\end{center}

Fig.1: Periodic strings: The string $S$ to the left has one attractive $\delta$-well 
with a single bound state, followed by a 
tunnel. The vertical bar to its left shows the energy interval 
for the band germ. 
This interval is 
the energy gauge $f(S)$ comprising the bound state. The string $S^2$ in the middle has two
attractive $\delta$-potentials and two bound states. The energy gauge 
$f(S^2)$ is 
unchanged but comprises two bound states. The same energy gauge $f(S^4)$ for
the string $S^4$ to the right comprises four bound states. 
\vspace{0.5cm}

A schematic view of the periodic scheme is given in Fig.1.
Now we can extend the analysis to $n \rightarrow \infty$:

${\bf 2 \, Prop}$: For an infinite periodic repetition of a fixed string, 
the energy gauge stays the same as for the initial string. The band 
germ generates a band. Within each band there is an 
{\em infinite set of  (pairs of) Bloch states whose energies} 
$E_K<0$ 
{\em fill up  the original band germ}.

\subsection{Bloch and bound states in a single band.}

We shall use the notation of \cite{KR1}. Consider the $1D$
Schrödinger stationary equation on the line $x \in {\bf R}$,
\begin{equation}
\label{gl1a}
\left[-\frac{\hbar^2}{2m} \frac{d^2}{dx^2} +V(x)\right]
\phi(x) = E \phi(x).
\end{equation}
with $V(x)$ a piecewise constant potential. Denote two
fundamental solutions and their derivatives by $\phi_1(x),\phi_1'(x)$,
$\phi_2(x),\phi_2'(x)$. Choose the initial values 
$\phi_1(0)=1,\,\phi_1'(0)=0,\; \phi_2(0)=0,\,\phi_2'(0)=1$
which implies a Wronskian equal to $1$.
The  $2 \times 2$ standard transfer matrix is defined as 
\begin{eqnarray}
\label{gl1b}
M(x) & = & 
\left[ 
\begin{array}{ll}
\phi_1(x) & \phi_2(x) \\
\phi_1'(x)& \phi_2'(x) \\
\end{array}
\right],
\\ \nonumber
M(0) & = & 1.
\end{eqnarray}
The transfer matrix obeys the first-order system of equations
which is equivalent to the Schrödinger equation.
A {\em second interpretation} of the transfer matrix,
which we shall adopt in what follows,  results because
the matrix $M(x)$ can also be shown to {\em propagate the 
fundamental system}.  

The Schrödinger equation for piecewise constant potentials
on a finite string can now be solved as follows \cite{KR3}: 
We first determine the transfer
matrix for simple building blocks of finite size at the fixed energy. Then 
we propagate the solutions over the full string by matrix multiplication.  
This matrix multiplication guarantees the continuity of the solutions
and of their derivatives. Specific boundary conditions can finally be
met by determining the appropriate linear combination of the 
two fundamental solutions.
For real matrices with Wronski determinant $1$ 
the transfer matrices belong to the matrix group $SL(2,R)$.
More details on this group and the isomorphic group $SU(1,1)$
can be found in \cite{KR1}. One should be careful in 
applying the group concepts to the transfer matrices, as can be seen
from the following transformation properties.

It will prove convenient to pass to a new fundamental systems
of solutions by choosing different initial values. With respect to 
the matrix formed by the solutions, this transformation is achieved by 
{\em right multiplication with a constant matrix},
\begin{equation}
\label{gl1c}
M(x) \rightarrow  M(x)C
\end{equation}
The {\em new transfer matrix which propagates the new system} 
$M(x)C$ is no longer
equal to $M(x)C$, instead it {\em is given by} $C^{-1}M(x)C$.
This propagating transfer matrix by itself does not admit an interpretation
in terms of fundamental solutions and their derivatives!
 
We now pass to a new system of fundamental solutions with different initial 
conditions  by right
multiplication with the matrix
\begin{equation}
\label{gl1}
\tilde{R}:= \sqrt{i}\, R =
\sqrt{\frac{1}{2\kappa}}
\left[ \begin{array}{ll}
1 & 1 \\
-\kappa & \kappa \\
\end{array}
\right].
\end{equation}
The system corresponding to the matrix $M(x)\tilde{R}$ has the initial 
data $\tilde{R}$ at $x=0$. At negative energy 
$E= -\frac{\hbar^2}{2m} \kappa^2$  we define a {\em tunnel} as a 
string with vanishing potential.
In a tunnel the transfer matrix takes the form
\begin{equation}
\label{gl2}
M(x)\tilde{R} = \sqrt{\frac{1}{2\kappa}} \left[
\begin{array}{ll}
\exp(-\kappa x)& \exp(\kappa x) \\
-\kappa \exp(-\kappa x)& \kappa \exp(\kappa x) \\
\end{array} \right].
\end{equation}
The transfer matrix which propagates 
this new system at negative energy is given by 
\begin{eqnarray}
\label{gl2a}
\tilde{M}(x) &:= & \tilde{R}^{-1}M(x)\tilde{R} 
\\ \nonumber
             & = & \left[ 
\begin{array}{ll}
\exp (-\kappa x) & 0 \\
0 & \exp (\kappa x) \\
\end{array} \right]
\end{eqnarray}
For a  bound state we require that an exponentially increasing
function at negative energy on the left-hand tunnel,
after passing an intermediate transfer matrix $\tilde{M}$,
produces an exponentially
decreasing function on the right-hand side. In the new basis, this property 
and the form eq. \ref{gl2} require that the intermediate transfer matrix
obeys $\tilde{M}_{22}=0$. 

${\bf 3\, Prop}$: In the system $\tilde{M}$ of transfer matrices
for finite strings, the
condition for a bound state requires that the second diagonal element
vanishes.
 
A tunnel of length $b$, followed by an attractive delta-potential of
strength $u$ at negative energy $E= -\frac{\hbar^2}{2m}
\kappa^2$ we term the string $S$. We define $\delta := u/\kappa$ 
and  $\lambda_1:=\exp(\beta),\; \beta := \kappa b$ to obtain the transfer 
matrix of $S$ as
\begin{eqnarray}
\label{gl3} 
\tilde{M}_1 & = & \left[
\begin{array}{ll}
(1+\frac{1}{2}\delta)& \frac{1}{2}\delta  \\
-\frac{1}{2}\delta & (1-\frac{1}{2}\delta)\\
\end{array} \right]
\left[
\begin{array}{ll}
\lambda_1^{-1}& 0 \\
0             & \lambda_1 \\
\end{array} \right]
\\ \nonumber
 & = & \left[
\begin{array}{ll}
\lambda_1^{-1}(1+\frac{1}{2}\delta)& \lambda_1\frac{1}{2}\delta  \\
-\lambda_1^{-1} \frac{1}{2}\delta & \lambda_1(1-\frac{1}{2}\delta)\\
\end{array} \right].
\end{eqnarray}
Here the first transfer matrix describes the delta-potential and the second
one the tunnel, compare \cite{KR3}.
This and all other  transfer matrices we take as a function of 
the dimensionless variable $\beta = \kappa b$, related to the 
energy and to the length of the cell $S$. We also consider
them as a function of the variable 
$\gamma = ub$ related to the strength of the delta-potentials, 
and study the family of systems with varying strength.
Instead of $\gamma$ we shall often  use the ratio $\delta := \frac{\gamma}{\beta}$
because then the  matrix elements of $\tilde{M}_1$
are (linear) {\em polynomials in the variable} $\delta$. Since we shall 
generate all other transfer matrices by matrix multiplication, a general 
transfer matrix will be a {\em polynomial in the parameter} $\delta$.
This will the basis for a {\em polynomial method} for handling the
analysis.
 
We introduce the following short-hand notation for the elements 
of a matrix $\tilde{M}_i$
\begin{eqnarray}
\label{gl4}
\tilde{M}_i & := & 
\left[
\begin{array}{ll}
a_i & b_i \\
c_i & d_i 
\end{array}
\right] 
\\ \nonumber
x_i& := &\frac{1}{2} (a_i+d_i)
, \\ \nonumber
y_i& := & \frac{1}{2} (a_i-d_i).
\end{eqnarray}
The second row of eq. \ref{gl4} determines the {\em half-trace} 
$x_i=\frac{1}{2} tr(\tilde{M})$.
For the matrix $\tilde{M}_1$ we find from eq. \ref{gl3}
\begin{eqnarray}
\label{gl5}
x_1& = & \frac{1}{2}(\lambda_1+\lambda_1^{-1}) 
-\frac{1}{4}\delta (\lambda_1-\lambda_1^{-1}),
\\ \nonumber
y_1& = & -\frac{1}{2}(\lambda_1-\lambda_1^{-1}) 
+\frac{1}{4}\delta (\lambda_1+\lambda_1^{-1}),\\ \nonumber
d_1& = & \lambda_1(1-\frac{1}{2}\delta).
\end{eqnarray}

\begin{center}
\begin{picture}(0,0)%
\includegraphics{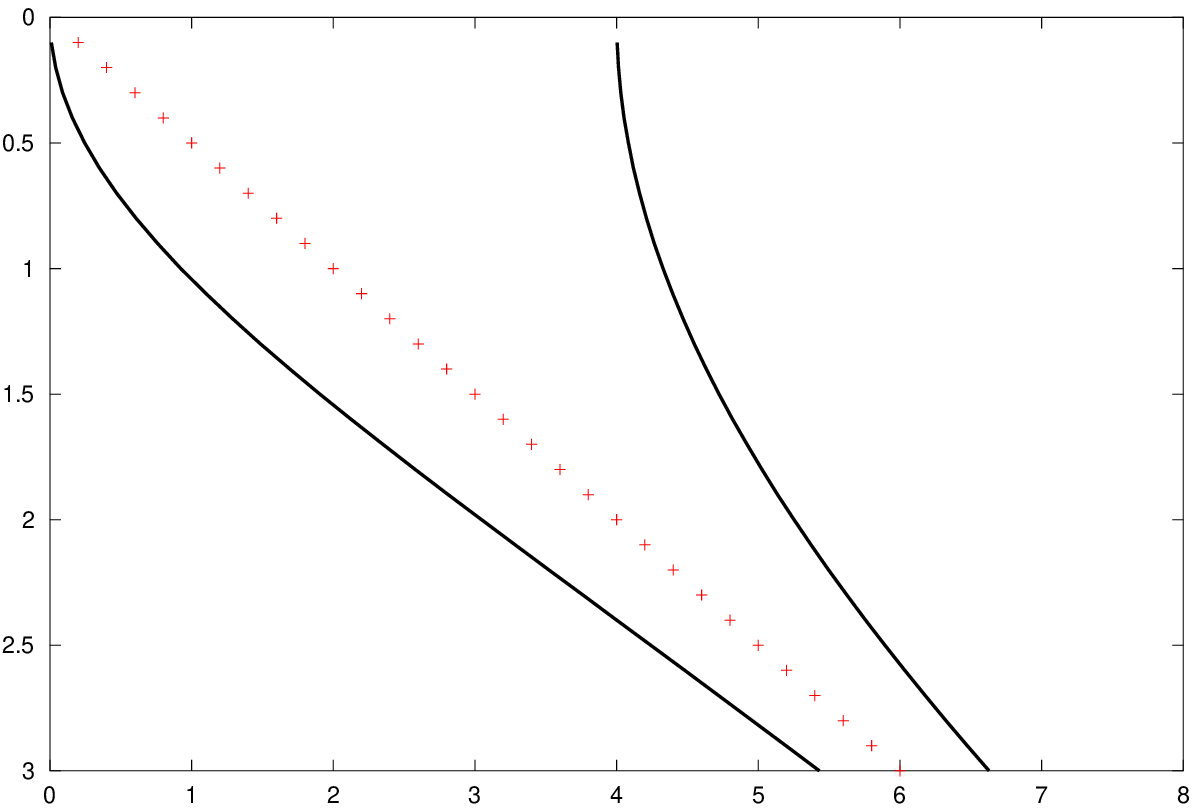}%
\end{picture}%
\setlength{\unitlength}{1973sp}%
\begingroup\makeatletter\ifx\SetFigFont\undefined
\def\x#1#2#3#4#5#6#7\relax{\def\x{#1#2#3#4#5#6}}%
\expandafter\x\fmtname xxxxxx\relax \def\y{splain}%
\ifx\x\y   
\gdef\SetFigFont#1#2#3{%
  \ifnum #1<17\tiny\else \ifnum #1<20\small\else
  \ifnum #1<24\normalsize\else \ifnum #1<29\large\else
  \ifnum #1<34\Large\else \ifnum #1<41\LARGE\else
     \huge\fi\fi\fi\fi\fi\fi
  \csname #3\endcsname}%
\else
\gdef\SetFigFont#1#2#3{\begingroup
  \count@#1\relax \ifnum 25<\count@\count@25\fi
  \def\x{\endgroup\@setsize\SetFigFont{#2pt}}%
  \expandafter\x
    \csname \romannumeral\the\count@ pt\expandafter\endcsname
    \csname @\romannumeral\the\count@ pt\endcsname
  \csname #3\endcsname}%
\fi
\fi\endgroup
\begin{picture}(11357,7615)(749,-8300)
\end{picture}

\end{center}

\begin{center}
\begin{picture}(0,0)%
\includegraphics{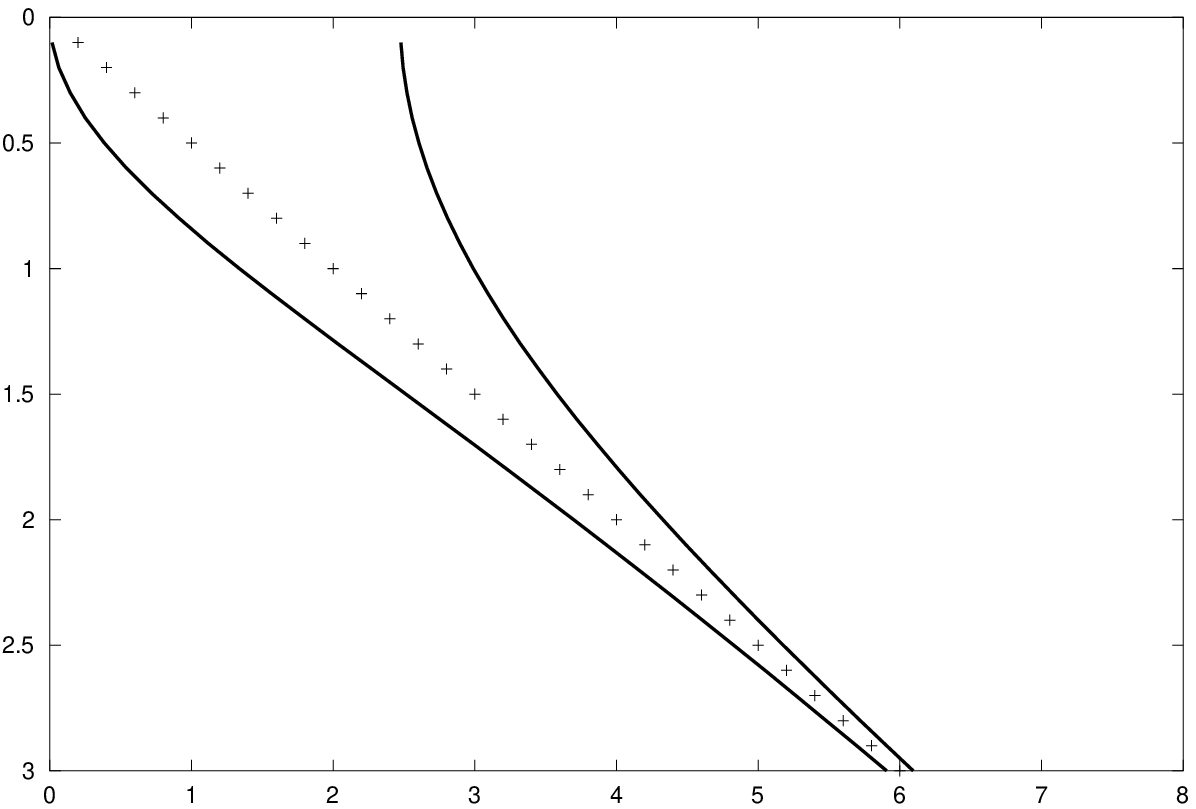}%
\end{picture}%
\setlength{\unitlength}{1973sp}%
\begingroup\makeatletter\ifx\SetFigFont\undefined
\def\x#1#2#3#4#5#6#7\relax{\def\x{#1#2#3#4#5#6}}%
\expandafter\x\fmtname xxxxxx\relax \def\y{splain}%
\ifx\x\y   
\gdef\SetFigFont#1#2#3{%
  \ifnum #1<17\tiny\else \ifnum #1<20\small\else
  \ifnum #1<24\normalsize\else \ifnum #1<29\large\else
  \ifnum #1<34\Large\else \ifnum #1<41\LARGE\else
     \huge\fi\fi\fi\fi\fi\fi
  \csname #3\endcsname}%
\else
\gdef\SetFigFont#1#2#3{\begingroup
  \count@#1\relax \ifnum 25<\count@\count@25\fi
  \def\x{\endgroup\@setsize\SetFigFont{#2pt}}%
  \expandafter\x
    \csname \romannumeral\the\count@ pt\expandafter\endcsname
    \csname @\romannumeral\the\count@ pt\endcsname
  \csname #3\endcsname}%
\fi
\fi\endgroup
\begin{picture}(11357,7600)(1048,-8297)
\end{picture}

\end{center}

Figs 2,3: The strings of length $\kappa b = 1,\tau$ respectively
produces two different band germs which encapsulate the same
bound state. Here and in all the corresponding Figures, 
the values of $\gamma$ and $\beta$ determine
the potential strength and the energy respectively and  are plotted 
as the horizontal and vertical coordinates respectively. 
\vspace{0.2cm}

We shall assume that we are in the band germ and, moreover,
that the upper edge of the band occurs at negative energy.
In sections $4,5$ we shall deal with more general scenarios. 
For $|x_1|\leq 1$, the eigenvalues of $\tilde{M}_1$ are complex conjugate
of modulus $1$ given by 
\begin{equation}
\label{gl6}
\theta_1= x_1 +i\sqrt{1-x_1^2},\; \theta_2= x_1 -i\sqrt{1-x_1^2}.
\end{equation}
So $x_i$ is the half-trace, and the local Bloch label $K$ for 
fixed energy is obtained from 
\begin{equation}
\label{gl7}
\cos(Kb)= x_1(\beta,\gamma),\; 0 \leq K \leq \frac{\pi}{b}.
\end{equation}

The values $\beta: x_1(\beta) = -1,1$ determine the two 
band edges respectively.

The lower diagonal element by $d_1(\beta)=0$ determines,
compare \cite{KR3} and what was explained above,   
the {\em single bound state of the delta-potential} at
$\delta = \frac{u}{\kappa}_0 =2$. The variable $\lambda_1$,
related to the tunnel length, drops out of this equation.

As $\tilde{M}$ is real, the eigenvectors can be chosen complex conjugate.
We find 
\begin{eqnarray}
\label{gl8}
\tilde{M}_1 V & = & V \Lambda, \\ \nonumber
V & = & \left[
\begin{array}{ll}
p & \overline{p} \\
v & \overline{v} \\
\end{array}\right] \\ \nonumber
\Lambda & = & \left[
\begin{array}{ll}
\theta_1 & 0 \\
0 & \theta_2 \\
\end{array} \right].
\end{eqnarray}
with
\begin{eqnarray}
\label{gl9}
\frac{v}{p} & = & \frac{\theta_1-a_1}{b_1} 
              =   \frac{-y_1+i\sqrt{1-x_1^2}}{b_1} 
\\ \nonumber
& = & \delta^{-1}
\left[(1-\lambda_1^{-2})-\frac{1}{2}\delta
(1+\lambda_1^{-2})+2i\lambda_1^{-1}\sqrt{1-x_1^2}\right].
\end{eqnarray}
With the help of the matrix $V$ we pass to the Bloch system
of solutions defined by $M(x)\tilde{R}V$ since it obeys 
$M(0)\tilde{R}V=\tilde{R}V,\; M(b)\tilde{R}V=\tilde{R}V\Lambda$. 
In line with eq. \ref{gl8} we mark derivatives with respect to
$x$ with a prime and put
\begin{eqnarray}
\label{gl10}
M_1(x)\tilde{R}V & := &
\left[
\begin{array}{ll}
\Phi_1& \Phi_2 \\
\Phi_1'&\Phi_2'
\end{array} \right],\\ \nonumber
\Phi_2& = &\overline{\Phi}_1,
\end{eqnarray}
and obtain for $0<x<b$ from eqs. \ref{gl8},\ref{gl9},\ref{gl10} 
the Bloch state
\begin{equation}
\label{gl11}
\Phi_1(x) := 
\frac{p}{\sqrt{2\kappa}} \left[\exp(-\kappa x)
+ \delta^{-1}
\left[(1-\lambda_1^{-2})-\frac{1}{2}\delta
(1+\lambda_1^{-2})+2i\lambda_1^{-1} \sqrt{1-x_1^2}\right]\exp(\kappa x)\right].
\end{equation}
To normalize the current density to the value 
$\frac{e\hbar}{2m}$ it suffices to  choose 
$det(V)= p\overline{v}-v\overline{p}=-i$ which together 
with eq. \ref{gl9} yields 
\begin{equation}
\label{gl12}
p\overline{p} = \frac{\delta\lambda_1}{4\sqrt{1-x_1^2}},\;
p  = i \frac{1}{2} \sqrt{\frac{\delta\lambda_1}{\sqrt{1-x_1^2}}},
\end{equation}
where we have chosen a particular phase for $p$ so that $\overline{p}=-p$.
The two Bloch states obey the bilinear relation
\begin{equation}
\label{gl13}
(-i)(\overline{\Phi_l}\Phi_j'-\overline{\Phi_l'}\Phi_j) 
=(-1)^{l+1}\delta_{lj}.
\end{equation}
which can be used to define a {\em scalar product} for Bloch states:

${\bf 4\, Def}$: To any pair $\Phi_1,\Phi_2$ of complex solutions 
of the Schr\"odinger equation we can associate an 
indefinite hermitian scalar product $\langle \Phi_1,\Phi_2 \rangle$ 
by the left-hand side of eq. \ref{gl13}.

The relation between the Bloch states and the 
exponential states $M_1(x)\tilde{R}$ is given from eqs.
\ref{gl8} and \ref{gl10} by
\begin{eqnarray}
\label{gl14}
\left[
\begin{array}{ll}
\psi_1& \psi_2 \\
\psi_1'&\psi_2'
\end{array} \right]
& := & M_1(x)\tilde{R} = (M_1(x) \tilde{R} V) \;V^{-1}
\\ \nonumber 
& = & 
\left[
\begin{array}{ll}
\Phi_1& \Phi_2 \\
\Phi_1'&\Phi_2'
\end{array} 
\right] i 
\left[
\begin{array}{rr}
\overline{v} & -\overline{p} \\
-v           & p
\end{array} \right]
\end{eqnarray}
A particular case arises if we choose the 
bound state energy $E = -\frac{\hbar^2}{2m} \kappa_0^2$ and 
construct the Bloch states eq. \ref{gl10}
for $\kappa=\kappa_0, \delta =2$.
This bound state must increase exponentially in the unit cell
towards the delta-potential and is given 
from eqs. \ref{gl11}, \ref{gl12} and \ref{gl14} 
for $0<x<b$ by 
\begin{equation}
\label{gl15a}
\psi_2 (x) = 
-\frac{1}{2} \sqrt{\frac{2 \lambda_1}{\sqrt{1-\lambda_1^{-2}}}}
(\Phi_1(x)+\Phi_2(x)).
\end{equation}
which differs from the expression given in \cite{KR3} eq.(22) in the phase
for the Bloch states and due to the new sequence of tunnel and 
potential.
A natural real unbound companion of this bound state from 
eq. \ref{gl14} is
\begin{equation}
\label{gl15b}
\psi_1 (x) = i (\Phi_1(x)\overline{v}-\Phi_2(x)v),
\end{equation}
because it can be easily be shown that the scalar product eq. \ref{gl13} 
for these two real states yields the Wronskian
\begin{equation}
\label{gl16}
\psi_1 \psi_2'-\psi_1'\psi_2 = 1.
\end{equation}
Here we wish to comment on the method of orthogonalized plane 
waves $OPW$ in which Bloch states are required to be orthogonal
to certain bound states \cite{AS} p.206-208. We doubt that this 
requirement can be handled with the standard integral scalar product
in Hilbert space,
or with a scalar product that involves only integration
over the unit cell: The first integration  does not apply 
without qualification to 
unbound states, and the second one introduces terms at the boundary
of the unit cell.

\subsection{The string $S^n$.}

Now we pass from the string $S$ by n-fold repetition to the string $S^n$
with the transfer matrix $\tilde{M}_1^n$. This power of $\tilde{M}_1$ can be
written as 
\begin{equation}
\label{gl17}
\tilde{M}_1^n = V \Lambda^n V^{-1}.  
\end{equation}
Explicitly the four matrix elements from eqs. \ref{gl8}, \ref{gl9}, 
\ref{gl10} are
\begin{eqnarray}
\label{gl18}
a_n & = & a(\tilde{M}_1^n)  =  \cos(nKb) + y_1 \frac{\sin(nKb)}{\sin(Kb)},
\\ \nonumber
d_n & = & d(\tilde{M}_1^n)  =  \cos(nKb) - y_1 \frac{\sin(nKb)}{\sin(Kb)},
\\ \nonumber
b_n & = & b(\tilde{M}_1^n)  =  \frac{1}{2} \lambda_1 \delta \frac{\sin(nKb)}{\sin(Kb)},
\\ \nonumber
c_n & = & c(\tilde{M}_1^n)  =  -\frac{1}{2} \lambda_1^{-1}\delta \frac{\sin(nKb)}{\sin(Kb)}.
\end{eqnarray} 
with $y_1$ taken from $\tilde{M}_1$ in eq. \ref{gl3}.

\subsection{Rational Bloch labels.}

Consider the transfer matrix eqs. \ref{gl17}, \ref{gl18} for the unit cell. 
For fixed integer $n>0$ at the points 
\begin{equation}
\label{gl19}
nKb= \mu \pi,\; \mu =0,\pm 1,\ldots ,\pm (n-1), \pm n
\end{equation}
it becomes
$\tilde{M}_1^n = (-1)^n e$. 
For any such Bloch label it follows that the Bloch states have the period 
$2nb$ with symmetry group $C_{2n}$. We can relate these states 
to a chain of $n$ atoms by imposing
as boundary conditions the factor $(-1)^n= \pm 1$ for even and odd $n$ 
respectively on the interval $nb$.
Moreover the Bloch state for fixed $\mu$ on the unit cell transforms according 
to the representation 
\begin{equation}
\label{gl20}
D^{\mu} = \exp (i \mu \frac{2\pi}{2n})
\end{equation}
for the generator of order $2n$ of the cyclic group $C_{2n}$.
 
There is {\em degeneracy} as we have a real hamiltonian: For opposite signs 
of $Kb$ and $\mu$ we have pairs of Bloch states with equal energy, running to the left and 
right respectively. The set of rational states for fixed $n$ yields a spectrum
of (in part) conjugate eigenstates. The energy increases with $|\mu|$.
The rational states are shown for $n=2,3$ in Figs. 5, 6.

With respect to $C_{2n}$, each eigenstate is 
characterized by a real or by a pair of complex conjugate irreducible
representations of $C_{2n}$ . All these rational states occur in the band and 
separate it into $n$ partial bands characterized by 
\begin{equation}
\label{gl20b}
K \geq 0:\; \frac{\mu \pi}{nb} \leq K  \leq \frac{(\mu+1) \pi}{nb},
\;  \mu =0, 1,\ldots ,(n-1), n .
\end{equation}
and similar expressions for $ K \leq 0$. 

\subsection{Bound states and clusters of the string $S^n$.}

The bound states of the finite string $S^n$ are characterized from eq. \ref{gl18}
by 
\begin{eqnarray}
\label{gl21}
d(\tilde{M}_1^n)& = & 0, \\ \nonumber
\tan(nKb) & = & \frac{\sin(Kb)}{y_1}.
\end{eqnarray}
The free string forms a cluster with $n$ atoms and hence at most $n$ bound states.
We now compare these bound states with the states of the (up to a sign)
periodic string of $n$ atoms:
In  the interval between two successive zeros 
of $\tan(nKB)$ with  $nKb =  \mu \pi, (\mu+1) \pi$, which correspond
to the rational $K$ labels associated with states of the periodic 
string, the value of $\tan(nKb)$ goes once to infinity.
The right-hand side of eq. \ref{gl21} 
is a smooth function of $\beta$ except
at the point $y_1(\beta)=0$. The left and right-hand side of eq. \ref{gl21}
are displayed  for the case $n=10$ as functions of $\beta$ in Fig.4.

\begin{center}
\begin{picture}(0,0)%
\includegraphics{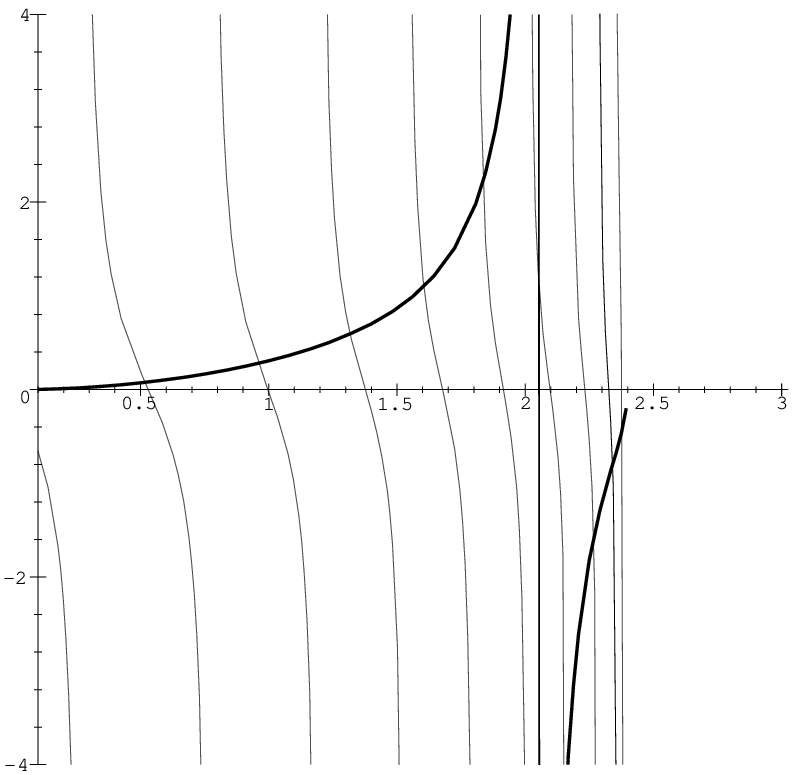}%
\end{picture}%
\setlength{\unitlength}{1973sp}%
\begingroup\makeatletter\ifx\SetFigFont\undefined
\def\x#1#2#3#4#5#6#7\relax{\def\x{#1#2#3#4#5#6}}%
\expandafter\x\fmtname xxxxxx\relax \def\y{splain}%
\ifx\x\y   
\gdef\SetFigFont#1#2#3{%
  \ifnum #1<17\tiny\else \ifnum #1<20\small\else
  \ifnum #1<24\normalsize\else \ifnum #1<29\large\else
  \ifnum #1<34\Large\else \ifnum #1<41\LARGE\else
     \huge\fi\fi\fi\fi\fi\fi
  \csname #3\endcsname}%
\else
\gdef\SetFigFont#1#2#3{\begingroup
  \count@#1\relax \ifnum 25<\count@\count@25\fi
  \def\x{\endgroup\@setsize\SetFigFont{#2pt}}%
  \expandafter\x
    \csname \romannumeral\the\count@ pt\expandafter\endcsname
    \csname @\romannumeral\the\count@ pt\endcsname
  \csname #3\endcsname}%
\fi
\fi\endgroup
\begin{picture}(7548,7317)(1165,-8545)
\put(8400,-4692){\makebox(0,0)[lb]{\smash{\SetFigFont{14}{16.8}{bf}$\beta$}}}
\end{picture}

\end{center}

Fig.4:  Left- and right-hand side of the binding eq. \ref{gl21} 
on a string of length $10$ for $\gamma = 4$ as a function of 
$\beta$. \vspace{0.4cm}
 
On any interval between consecutive zeros of $\tan(nKb)$ 
there is precisely one intersection point of the left and the right-hand expression
of eq. \ref{gl21}.

In terms of boundary conditions this implies that, in between 
the energies of two
consecutive rational values of $Kb$ with periodic boundary conditions,
there is precisely one bound state of the $n$-atom cluster. 
This state, termed a 
{\em decaying state}
in \cite{KR3}, is a single linear
combination of two degenerate Bloch states on the interval of length 
$nb$ and, when the $n$-atom string is cut out, admits  exponential decay 
to the right and to the left. 

${\bf 5\, Prop}$: Any set of rational Bloch labels eq. \ref{gl19} separates 
the band into $n$ partial bands. Each partial band contains a 
single bound state
of the free $n$-atom string.

The relation between rational Bloch labels and bound states is 
shown in the following Figs. 5,6.

\begin{center}
\begin{picture}(0,0)%
\includegraphics{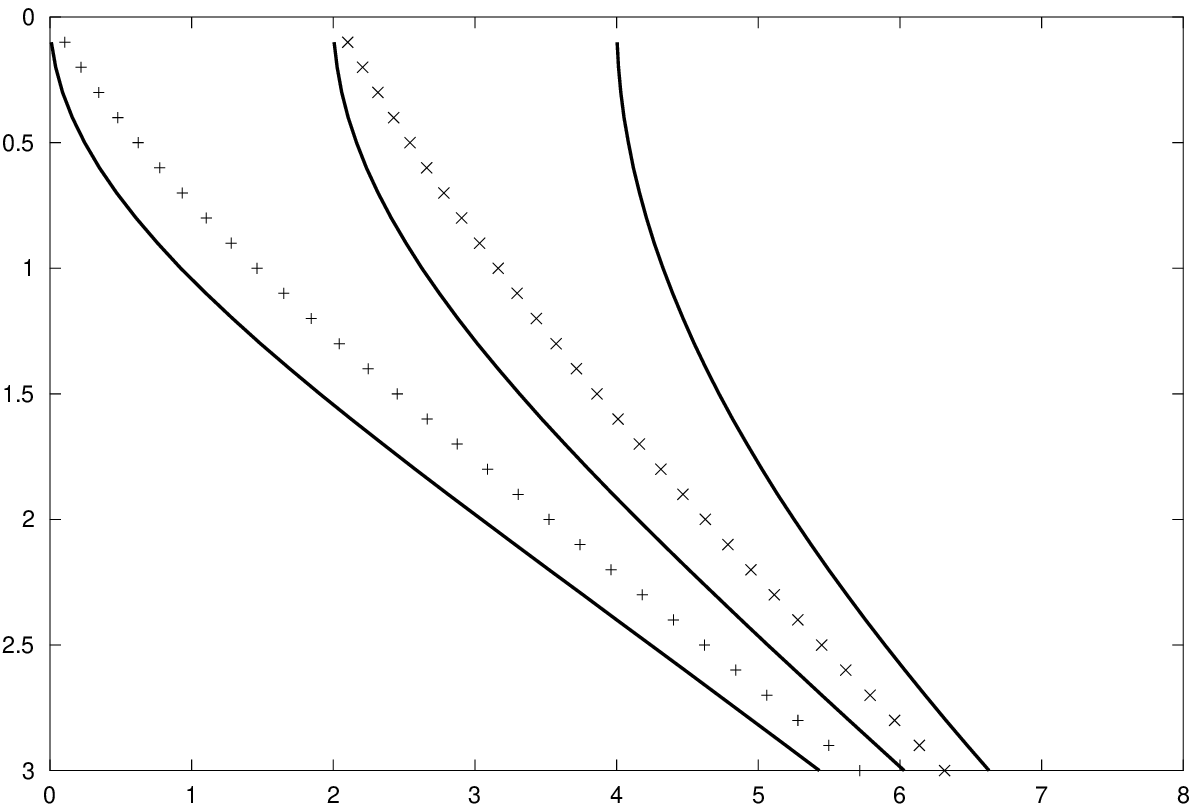}%
\end{picture}%
\setlength{\unitlength}{1973sp}%
\begingroup\makeatletter\ifx\SetFigFont\undefined
\def\x#1#2#3#4#5#6#7\relax{\def\x{#1#2#3#4#5#6}}%
\expandafter\x\fmtname xxxxxx\relax \def\y{splain}%
\ifx\x\y   
\gdef\SetFigFont#1#2#3{%
  \ifnum #1<17\tiny\else \ifnum #1<20\small\else
  \ifnum #1<24\normalsize\else \ifnum #1<29\large\else
  \ifnum #1<34\Large\else \ifnum #1<41\LARGE\else
     \huge\fi\fi\fi\fi\fi\fi
  \csname #3\endcsname}%
\else
\gdef\SetFigFont#1#2#3{\begingroup
  \count@#1\relax \ifnum 25<\count@\count@25\fi
  \def\x{\endgroup\@setsize\SetFigFont{#2pt}}%
  \expandafter\x
    \csname \romannumeral\the\count@ pt\expandafter\endcsname
    \csname @\romannumeral\the\count@ pt\endcsname
  \csname #3\endcsname}%
\fi
\fi\endgroup
\begin{picture}(11357,7600)(749,-8747)
\end{picture}

\end{center}

\begin{center}
\begin{picture}(0,0)%
\includegraphics{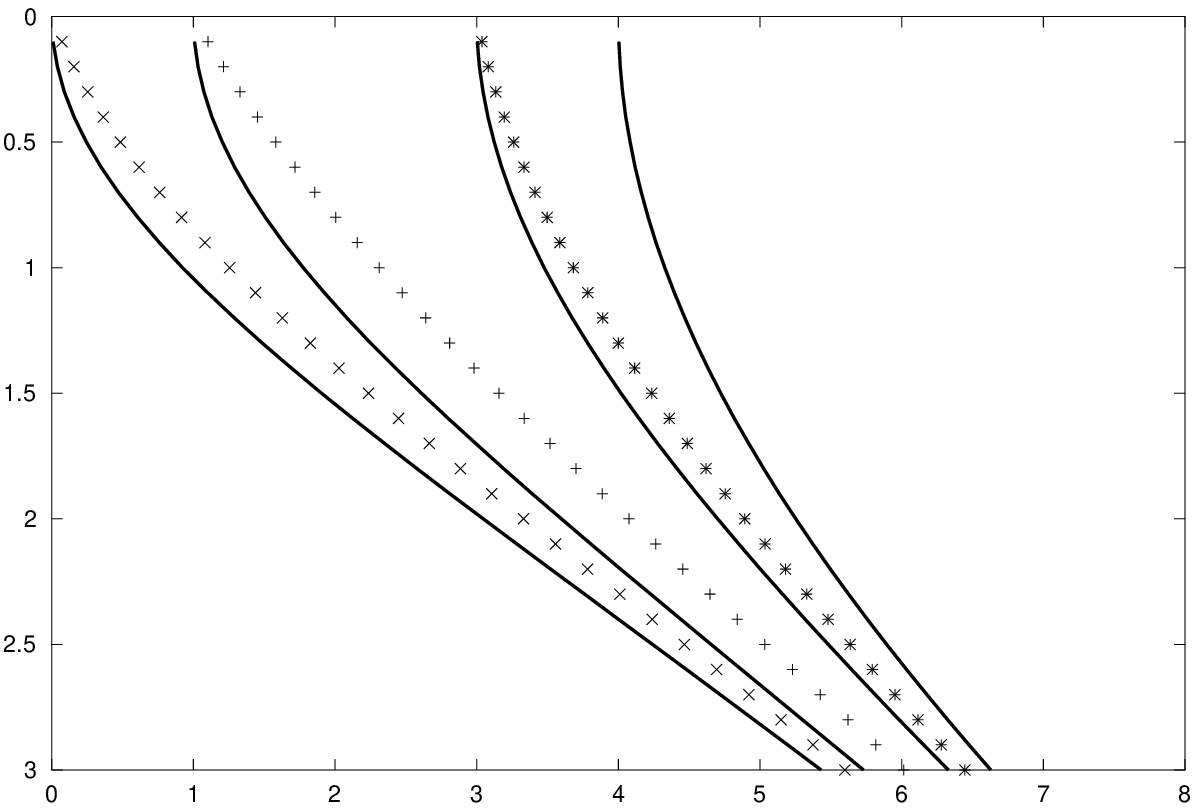}%
\end{picture}%
\setlength{\unitlength}{1973sp}%
\begingroup\makeatletter\ifx\SetFigFont\undefined
\def\x#1#2#3#4#5#6#7\relax{\def\x{#1#2#3#4#5#6}}%
\expandafter\x\fmtname xxxxxx\relax \def\y{splain}%
\ifx\x\y   
\gdef\SetFigFont#1#2#3{%
  \ifnum #1<17\tiny\else \ifnum #1<20\small\else
  \ifnum #1<24\normalsize\else \ifnum #1<29\large\else
  \ifnum #1<34\Large\else \ifnum #1<41\LARGE\else
     \huge\fi\fi\fi\fi\fi\fi
  \csname #3\endcsname}%
\else
\gdef\SetFigFont#1#2#3{\begingroup
  \count@#1\relax \ifnum 25<\count@\count@25\fi
  \def\x{\endgroup\@setsize\SetFigFont{#2pt}}%
  \expandafter\x
    \csname \romannumeral\the\count@ pt\expandafter\endcsname
    \csname @\romannumeral\the\count@ pt\endcsname
  \csname #3\endcsname}%
\fi
\fi\endgroup
\begin{picture}(11357,7750)(1499,-9676)
\end{picture}

\end{center}

Figs. 5,6: The strings $S^2$ and $S^3$. In the supercell analysis,
two or three band germs are glued together without gaps and overlaps.
Each band includes a single bound or decaying state. The horizontal 
coordinate is the potential strength $\gamma$, the vertical 
coordinate the energy $\beta$. Notice the relation between the onset
of additional bound states and the subband edges.
\vspace{0.2cm}

Since the rational labels $K$ form a dense set of energies, 
we must also have
a dense set of bound or decaying  state energies of free 
$n$-atom strings. Each 
bound state  coincides on a cell of length $nb$ with  a linear superposition 
of two 
Bloch states at the same fixed negative energy  similar to 
$\psi_2$ in eq. \ref{gl15a}.
Each one has a companion state of a form similar to 
$\psi_1$ in eq. \ref{gl15b}, and these two states have a Wronskian as
in eq. \ref{gl16}.

\subsection{Participation number.}

From the relation between $n$-atom clusters and periodic states we can 
assign within a band a  {\em participation number} $n$
to the rational Bloch vectors of the form $K=\frac{\mu\pi}{nb}$,
eq. \ref{gl19}: This 
number determines the number $n$ of atoms which participate in the Bloch 
states whose period up to a sign is $nb$, and at the same time 
assigns $n$  bound or decaying states of the 
cluster. Small values of $n$ indicate states which involve a few
atoms and may be called localized with respect to the clusters.

\subsection{Supercell interpretation.}

An alternative interpretation of the results given above can be given 
in terms of a {\em supercell scheme}: We consider 
the supercell of length $nb$ formed from $n$ centers. 
The corresponding Brillouin zones have the
size $2\pi/(2bn)$. The positive Bloch labels for the $n$ bands in 
this scheme are obtained from eq. \ref{gl20b} as 
\begin{eqnarray}
\label{gl24}
0 \leq K(n,\mu ) & = & K - \frac{\mu \pi}{nb} \leq \frac{\pi}{nb},
\\ \nonumber
\mu & = & 0, 1,\ldots , (n-1), n
\end{eqnarray}
and similar expressions for $K(n,\mu) \leq 0.$
With the new Bloch label $K(n,\mu)$, these  superbands are 
identical with the partial bands for the rational values of $K$.

${\bf 6\, Prop}$: The superbands for the supercell of length $nb$ are glued 
together without gaps and overlaps and fill up the single band. Their
Bloch states are identical to those of the  
partial bands of the rational scheme.
Within each superband there is included a single bound state. This
bound state coincides on the supercell with a bound state of the 
free n-atom cluster which forms the motive of
the supercell. The alternating pattern of bound and periodic $n$-atom
states illuminates the structure of the band.

\subsection{Large $n$ limit.}

Now we consider the limit $n \rightarrow \infty $. 
The rational $K$-labels form a dense
set on the Brillouin zone. The length of the periodic chains and of the 
clusters increases with the participation number. As the rational 
numbers form a dense set on the Brillouin zone, we have pairs of
Bloch states and bound states near any value of $K$ in the band. By picking  
rational values of $K$ characterized by $n$, we select a set of periodic states 
which pairwise encapsulate bound states of the finite string $S^n$.

From the equal  
spacing of the zeros of $\tan(nKb)$ with respect to the band label $K$,
and from the inclusion property for bound states there follows

${\bf 7\, Prop}$:  The density of bound and of Bloch states in the limit 
$n \rightarrow \infty$ is 
a constant function of the Bloch label $K$. In this limit, the 
density of (bound) states per unit energy interval fulfills the well-known 
relation
\begin{equation}
\label{gl25}
\frac{dN}{dE} \sim  \frac{dK}{dE}.
\end{equation} 

Although this is a smooth function in the limit, its rational 
approximants with large but finite $n$ may look quite irregular
as functions of $\beta$.

\section{Finite Quasiperiodic strings at negative energy.}

As an example of a quasiperiodic system we shall take the well-known
Fibonacci system: We maintain delta-potentials of equal strength 
for the atoms
but admit two different intervals $S$ and  $L$ of length $b$ and $qb$ 
respectively. For the proper Fibonacci case we choose 
$q=\tau =\frac{1}{2}(1+\sqrt{5})$, but for comparison with the 
periodic case we also consider the value $q=1$.  

The Fibonacci system we define algebraically by the recursive words $W_m$ and initial data
in the alphabet $\langle S,L \rangle$,
\begin{equation}
\label{gl26}
W_{m+1}:= W_{m-1}W_m,\; W_1=S,W_2=L.
\end{equation}
The word $W_m$ contains $f_m$ letters, $f_{m-1}$ letters $S$, and $f_{m}$ 
letters $L$. Here $f_m$ are the integer Fibonacci numbers defined by 
$ f_{m+1}=f_{m-1}+f_m,\; f_1=f_2=1$.

\subsection{Preview: Energy gauge in Fibonacci strings.}

As an introduction we take from \cite{KR3} the example 
of two strings $S,L$ which represent 
the same attractive delta-potentials combined with tunnels of
length $b$ and $\tau b$ respectively and then pass to the
string $SL$. The energy gauges for $S$ and $L$ both 
contain the same bound state. The string $SL$ has two bound states
but may have, depending on the tunnel length, two or one band germ.
In the first case there exists a region of  negative energy where
the single strings admit band germs but the combined string 
$SL$ does not. This case is schematically represented in Fig.7.

\begin{center}
\begin{picture}(0,0)%
\includegraphics{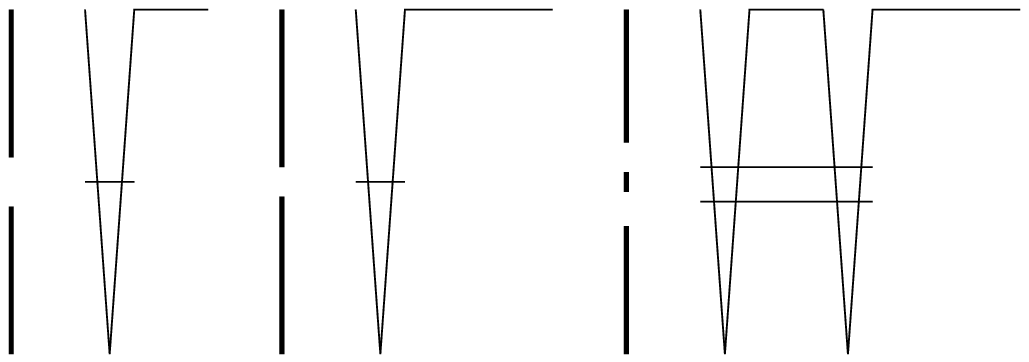}%
\end{picture}%
\setlength{\unitlength}{2072sp}%
\begingroup\makeatletter\ifx\SetFigFont\undefined
\def\x#1#2#3#4#5#6#7\relax{\def\x{#1#2#3#4#5#6}}%
\expandafter\x\fmtname xxxxxx\relax \def\y{splain}%
\ifx\x\y   
\gdef\SetFigFont#1#2#3{%
  \ifnum #1<17\tiny\else \ifnum #1<20\small\else
  \ifnum #1<24\normalsize\else \ifnum #1<29\large\else
  \ifnum #1<34\Large\else \ifnum #1<41\LARGE\else
     \huge\fi\fi\fi\fi\fi\fi
  \csname #3\endcsname}%
\else
\gdef\SetFigFont#1#2#3{\begingroup
  \count@#1\relax \ifnum 25<\count@\count@25\fi
  \def\x{\endgroup\@setsize\SetFigFont{#2pt}}%
  \expandafter\x
    \csname \romannumeral\the\count@ pt\expandafter\endcsname
    \csname @\romannumeral\the\count@ pt\endcsname
  \csname #3\endcsname}%
\fi
\fi\endgroup
\begin{picture}(9472,3800)(451,-3705)
\put(1351,-61){\makebox(0,0)[lb]{\smash{\SetFigFont{12}{14.4}{rm}$S$}}}
\put(2926,-61){\makebox(0,0)[lb]{\smash{\SetFigFont{12}{14.4}{rm}$f(L)$}}}
\put(3826,-61){\makebox(0,0)[lb]{\smash{\SetFigFont{12}{14.4}{rm}$L$}}}
\put(451,-61){\makebox(0,0)[lb]{\smash{\SetFigFont{12}{14.4}{rm}$f(S)$}}}
\put(5851,-61){\makebox(0,0)[lb]{\smash{\SetFigFont{12}{14.4}{rm}$f(SL)$}}}
\put(6976,-61){\makebox(0,0)[lb]{\smash{\SetFigFont{12}{14.4}{rm}$SL$}}}
\end{picture}

\end{center}

Fig.7: Fibonacci strings: The string $S$ to the left and its energy
gauge $f(S)$ are as in Fig. 1. The string $L$ in the middle has the same 
bound state within the gauge $f(L)$ of smaller width. 
The string $SL$ to the right has two attractive $\delta$-potentials and
hence the same bound states as $S^2$ in Fig. 1. In contrast to $S^2$,
the string $SL$ has  two separate band germs. 
The corresponding two parts of 
the energy gauge $f(SL)$ comprise each one bound state but block the energy of the
single-atom bound state.\vspace{0.5cm}

\subsection{Substitional systems and their invariants.}

The Fibonacci system to be considered can be seen as a particular case 
of a substitution system. We shall restrict our attention to 
substitutional systems generated by automorphisms of the free
group. For these algebraic concepts in general we refer to \cite{KR4}
and follow in detail \cite{KR1}. 
Consider the free group $F_2$ with generators (alphabet) $a_1,a_2$. Its elements
are all the words in the generators and their inverses. An automorphism
of $F_2$ is defined as a map 
\begin{equation}
\label{gl26a}
\rho: a_1,a_2 \rightarrow f_1(a_1,a_2),f_2(a_1,a_2)
\end{equation}
which is {\em algebraically invertible} so that $a_1,a_2$ can be expressed 
in terms of $f_1,f_2$. 

{\em Example}: Consider the substitution 
$a_1,a_2 \rightarrow f_1=a_2, f_2=a_1a_2$. Its inverse is given by
$a_1=f_2f_1^{-1},\; a_2=f_1$. This substitution describes in abstract 
terms the Fibonacci systems to be discussed below.

For any automorphism of $F_2$ we have the general 
theorem 

${\bf 8\, Prop}$ (Nielsen 1918): For $\rho$ an automorphism of $F_2$,
the commutator ${\cal K}(a_1,a_2):= a_1a_2a_1^{-1}a_2^{-1}$ obeys
\begin{equation}
\label{gl26b}
\rho ({\cal K}) = w {\cal K}^{\pm1}w^{-1}
\end{equation}
with $w \in F_2$.

{\em Example}: For the Fibonacci automorphism one finds
\begin{equation}
\label{gl26c}
\rho({\cal K}) = {\cal K}^{-1}
\end{equation}

From this abstract algebraic set-up we can pass to matrix systems by
first mapping the generators $a_1,a_2$ to two elements of a
matrix group like $SL(2,R)$ for the transfer matrices and then
performing the automorphism $\rho$ induced on the matrix group.
For certain classes of such induced automorphisms, 
one can obtain the (half-) traces of the images directly 
from the (half-) traces of the generators. This question is
studied in the theory of {\em trace maps} for which we refer to
Peyri\`{e}re \cite{PE}.
The Fibonacci automorphism and related systems belong
to this class \cite{KR1}. The Nielsen theorem in connection with
the induced automorphisms now provides one or more 
{\em invariants of induced automorphisms}: Consider the
induced commutator ${\cal K}$ on the matrix group $SL(2,R)$ and take its
trace. Under the Fibonacci automorphism the commutator as a matrix
is transformed into its inverse. Since all matrices are unimodular,
we get on the matrix group 
\begin{equation}
\label{gl26e}
tr(\rho({\cal K}))= tr({\cal K}^{-1})= tr({\cal K})
\end{equation}
${\bf 9\, Prop}$: The trace of the commutator is an invariant 
under the Fibonacci substitution.

The commutator $K(g_1,g_2)$ in a matrix group is a measure of 
the non-commutativity. In particular if the two matrices commute,
the half-trace eq. \ref{gl26e} must be $1$. Conversely if ${\cal K}$ equals the unit matrix,
the elements $g_1,g_2$ commute. We shall come back to this 
property in later sections.

\subsection{Recursive calculation of the transfer matrix.} 

To represent the Fibonacci  system by delta-potentials and the states by
transfer matrices we define 
\begin{equation}
\label{gl27}
\tilde{M}(S) := \tilde{M}_1,\; \tilde{M}(L) := \tilde{M}_2.
\end{equation}
We represent the Fibonacci string $W_{m+1}$ of eq. \ref{gl26} 
by the transfer matrix
\begin{equation}
\label{gl28}
\tilde{M}_{m+1}= \tilde{M}_{m-1} \tilde{M}_m.
\end{equation}
The transfer matrix $\tilde{M}_2$ has the same analytic form as
$\tilde{M}_1$ given in eq. \ref{gl3} with the same variable $\delta$ 
and the same energy $E$ but
with the replacement 
\begin{equation}
\label{gl29}
\lambda_1= \exp(\beta) \rightarrow 
\lambda_2 :=\exp(q\beta).
\end{equation}
Similar systems have been studied extensively in the literature,
compare references given by Kohmoto \cite{KOH} and in \cite{BAA}.
Many of these studies used a discrete version or
examined the recursion numerically. In what follows we shall use
the recursive method to explore the {\em analytic structure} of the
Fibonacci strings as functions of the variables $\beta,\delta$. 
We recall that $\tilde{M}_1,\tilde{M}_2$ eqs. \ref{gl3}, \ref{gl27} 
are linear polynomials
with respect to $\delta$. From this property and from eq. \ref{gl28} 
we get immediately :

${\bf 10\,Prop}$:
The matrix elements of $\tilde{M}_m$
are polynomials of degree $f_m$ with respect to the parameter $\delta$, 
with coefficients which are functions of $\beta$ via the expressions 
$\lambda_1,\lambda_2$.
  
For the finite Fibonacci string we wish to study the matrix elements
of the transfer matrix $\tilde{M}_i$ and in particular their combinations
$x_i,y_i,d_i=(x_i-y_i)$, since they yield information on the eigenvalues
and bound states. Two methods are available for the computation:

(i) We can use the full matrix recursion eq. \ref{gl28}, with the starting 
matrices $\tilde{M}_1,\tilde{M}_2$.
 
(ii)  We can use recursion
techniques for the half-traces and the other matrix elements,
as was proposed in \cite{KR1}. From this reference we deduce the following
recursion equations for the combinations of matrix
elements:
\begin{eqnarray}
\label{gl30}
\left[
\begin{array}{ll}
a_{m+1} & b_{m+1} \\
c_{m+1} & d_{m+1}
\end{array}
\right]
& = & 
\left[
\begin{array}{ll}
a_m+d_m & 0 \\
0 & a_m+d_m
\end{array}
\right]
\left[
\begin{array}{ll}
a_{m-1} & b_{m-1} \\
c_{m-1} & d_{m-1}
\end{array}
\right] \\ \nonumber
& - & 
\left[
\begin{array}{ll}
d_{m-2} & -b_{m-2} \\
-c_{m-2} & a_{m-2}
\end{array}
\right],
\\ \nonumber
x_{m+1} & = & 2x_mx_{m-1} -x_{m-2}, 
\\ \nonumber 
y_{m+1} & = & 2x_my_{m-1} +y_{m-2}, 
\\ \nonumber 
d_{m+1} & = & x_{m+1}-y_{m+1}.
\end{eqnarray}
To start these recursive relations we need the matrices 
$\tilde{M}_1,\tilde{M}_2$ given from eqs. \ref{gl3}, \ref{gl27} and 
compute $\tilde{M}_3$ from 
\begin{eqnarray}
\label{gl31a}
\tilde{M}_3 & = & \tilde{M}_1\tilde{M}_2: \\ \nonumber
a_3 & = & \lambda_1^{-1}\lambda_2^{-1}(1+\frac{1}{2}\delta)^2
        -\frac{1}{4}\lambda_1\lambda_2^{-1} \delta^2,
\label{gl31} \\ \nonumber
b_3 & = & \frac{1}{2}\lambda_1^{-1}\lambda_2\delta(1+\frac{1}{2}\delta)
        + \frac{1}{2}\lambda_1\lambda_2^{-1}\delta(1-\frac{1}{2}\delta)
\\ \nonumber
c_3 & = & -\frac{1}{2}\lambda_1^{-1}\lambda_2^{-1}\delta(1+\frac{1}{2}\delta)
          -\frac{1}{2}\lambda_1\lambda_2^{-1}\delta(1-\frac{1}{2}\delta)
\\ \nonumber
d_3 & = & \lambda_1\lambda_2(1-\frac{1}{2}\delta)^2
        -\frac{1}{4}\lambda_1^{-1}\lambda_2 \delta^2.
\end{eqnarray}
These matrix elements are polynomials of degree $2$ in the variable
$\delta$ as expected. The half-traces $x_m$ form a recursive system
by themselves. This is the well-known {\em Fibonacci trace map}.

${\bf 11\,Prop}$: By use of the recursive relations eq. \ref{gl30} 
we can construct algebraic polynomial 
expressions of degree $f_m$  for the half-trace $x_m(\beta,\delta)$ and for the 
matrix element $d_m(\beta,\delta)$, which we call the {\em band polynomial} and
the {\em bound polynomial} respectively. 
The edges of the band germs for the
Fibonacci string $W_m$  are given by the $f_m$ {\em roots of the 
band polynomial 
equations}
\begin{equation}
\label{gl32}
x_m(\beta,\delta)= \pm 1,
\end{equation}
while the bound states are given as the $f_m$ {\em roots of the bound polynomial 
equations} 
\begin{equation}
d_m(\beta,\delta)= 0.
\label{gl33}
\end{equation}
The $f_m$ roots of the polynomials eq. \ref{gl32} are real 
since they represent special
half-traces of real matrices. The $f_m$ roots of the polynomial eq. \ref{gl33} 
are real
for sufficient strength of the delta-potential since then we must get 
$f_m$ bound states.

For low values of $m$ the roots of the polynomials can be obtained in closed 
algebraic form.

\begin{center}
\begin{picture}(0,0)%
\includegraphics{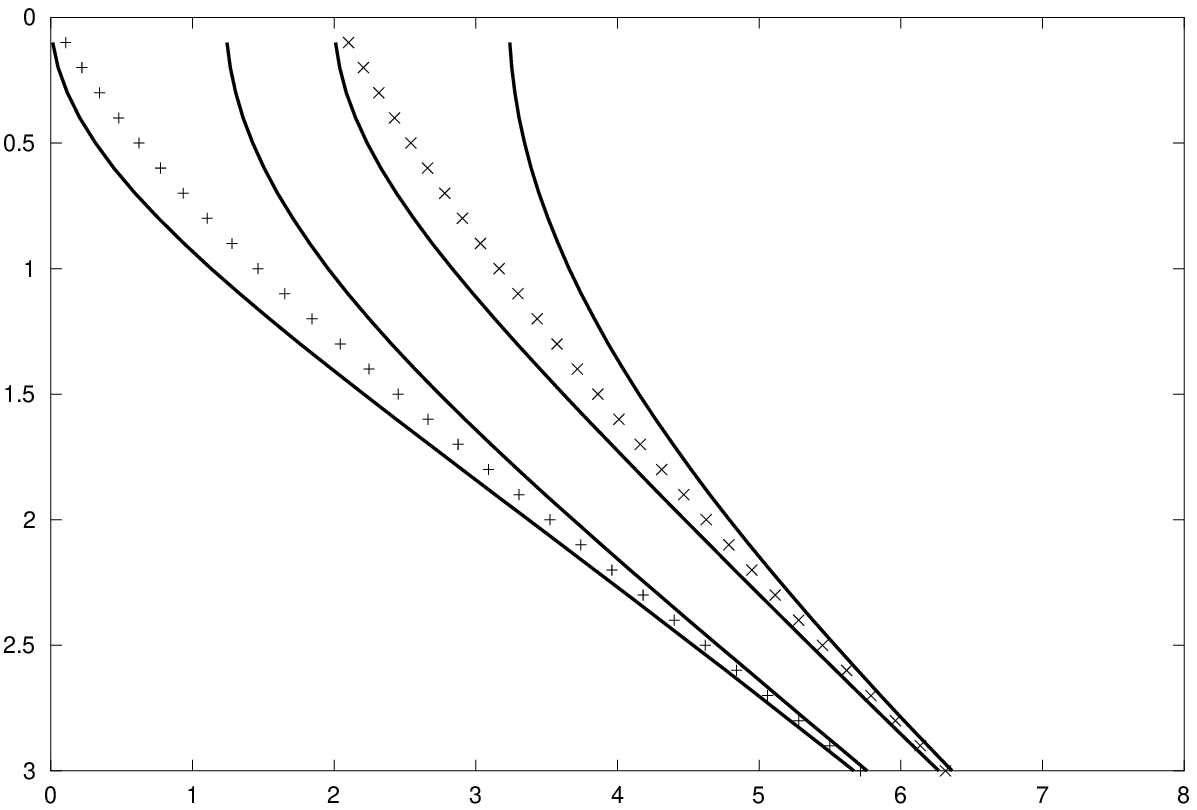}%
\end{picture}%
\setlength{\unitlength}{1973sp}%
\begingroup\makeatletter\ifx\SetFigFont\undefined
\def\x#1#2#3#4#5#6#7\relax{\def\x{#1#2#3#4#5#6}}%
\expandafter\x\fmtname xxxxxx\relax \def\y{splain}%
\ifx\x\y   
\gdef\SetFigFont#1#2#3{%
  \ifnum #1<17\tiny\else \ifnum #1<20\small\else
  \ifnum #1<24\normalsize\else \ifnum #1<29\large\else
  \ifnum #1<34\Large\else \ifnum #1<41\LARGE\else
     \huge\fi\fi\fi\fi\fi\fi
  \csname #3\endcsname}%
\else
\gdef\SetFigFont#1#2#3{\begingroup
  \count@#1\relax \ifnum 25<\count@\count@25\fi
  \def\x{\endgroup\@setsize\SetFigFont{#2pt}}%
  \expandafter\x
    \csname \romannumeral\the\count@ pt\expandafter\endcsname
    \csname @\romannumeral\the\count@ pt\endcsname
  \csname #3\endcsname}%
\fi
\fi\endgroup
\begin{picture}(11357,7795)(824,-8560)
\end{picture}

\end{center}

\begin{center}
\begin{picture}(0,0)%
\includegraphics{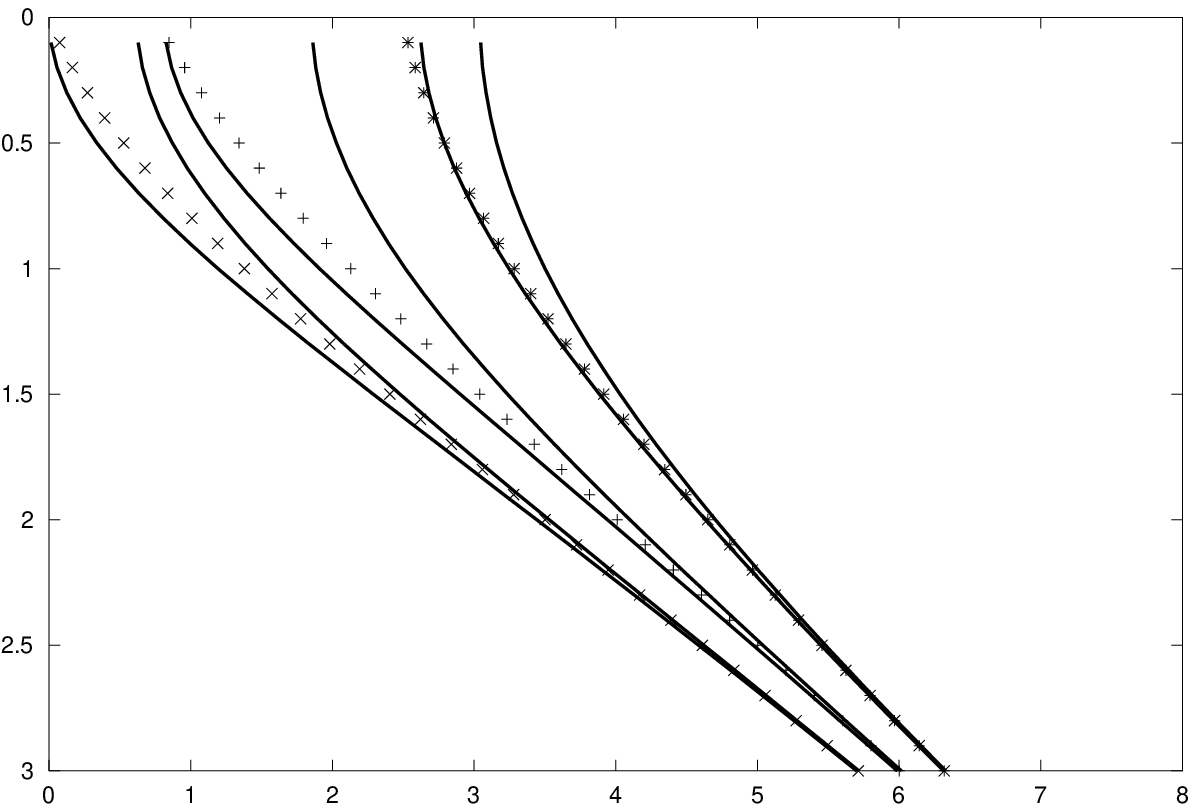}%
\end{picture}%
\setlength{\unitlength}{1973sp}%
\begingroup\makeatletter\ifx\SetFigFont\undefined
\def\x#1#2#3#4#5#6#7\relax{\def\x{#1#2#3#4#5#6}}%
\expandafter\x\fmtname xxxxxx\relax \def\y{splain}%
\ifx\x\y   
\gdef\SetFigFont#1#2#3{%
  \ifnum #1<17\tiny\else \ifnum #1<20\small\else
  \ifnum #1<24\normalsize\else \ifnum #1<29\large\else
  \ifnum #1<34\Large\else \ifnum #1<41\LARGE\else
     \huge\fi\fi\fi\fi\fi\fi
  \csname #3\endcsname}%
\else
\gdef\SetFigFont#1#2#3{\begingroup
  \count@#1\relax \ifnum 25<\count@\count@25\fi
  \def\x{\endgroup\@setsize\SetFigFont{#2pt}}%
  \expandafter\x
    \csname \romannumeral\the\count@ pt\expandafter\endcsname
    \csname @\romannumeral\the\count@ pt\endcsname
  \csname #3\endcsname}%
\fi
\fi\endgroup
\begin{picture}(11357,7795)(674,-8860)
\end{picture}

\end{center}

\begin{center}
\begin{picture}(0,0)%
\includegraphics{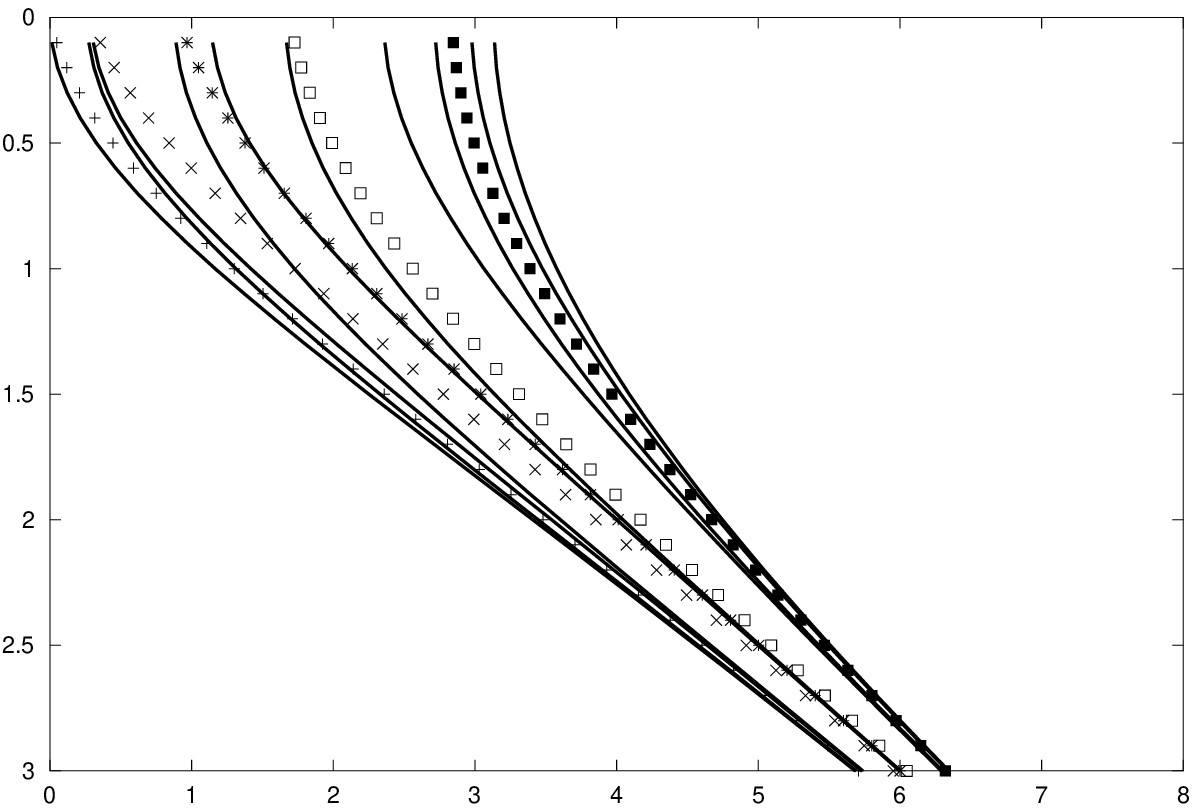}%
\end{picture}%
\setlength{\unitlength}{1973sp}%
\begingroup\makeatletter\ifx\SetFigFont\undefined
\def\x#1#2#3#4#5#6#7\relax{\def\x{#1#2#3#4#5#6}}%
\expandafter\x\fmtname xxxxxx\relax \def\y{splain}%
\ifx\x\y   
\gdef\SetFigFont#1#2#3{%
  \ifnum #1<17\tiny\else \ifnum #1<20\small\else
  \ifnum #1<24\normalsize\else \ifnum #1<29\large\else
  \ifnum #1<34\Large\else \ifnum #1<41\LARGE\else
     \huge\fi\fi\fi\fi\fi\fi
  \csname #3\endcsname}%
\else
\gdef\SetFigFont#1#2#3{\begingroup
  \count@#1\relax \ifnum 25<\count@\count@25\fi
  \def\x{\endgroup\@setsize\SetFigFont{#2pt}}%
  \expandafter\x
    \csname \romannumeral\the\count@ pt\expandafter\endcsname
    \csname @\romannumeral\the\count@ pt\endcsname
  \csname #3\endcsname}%
\fi
\fi\endgroup
\begin{picture}(11357,7990)(749,-8975)
\end{picture}

\end{center}

\begin{center}
\begin{picture}(0,0)%
\includegraphics{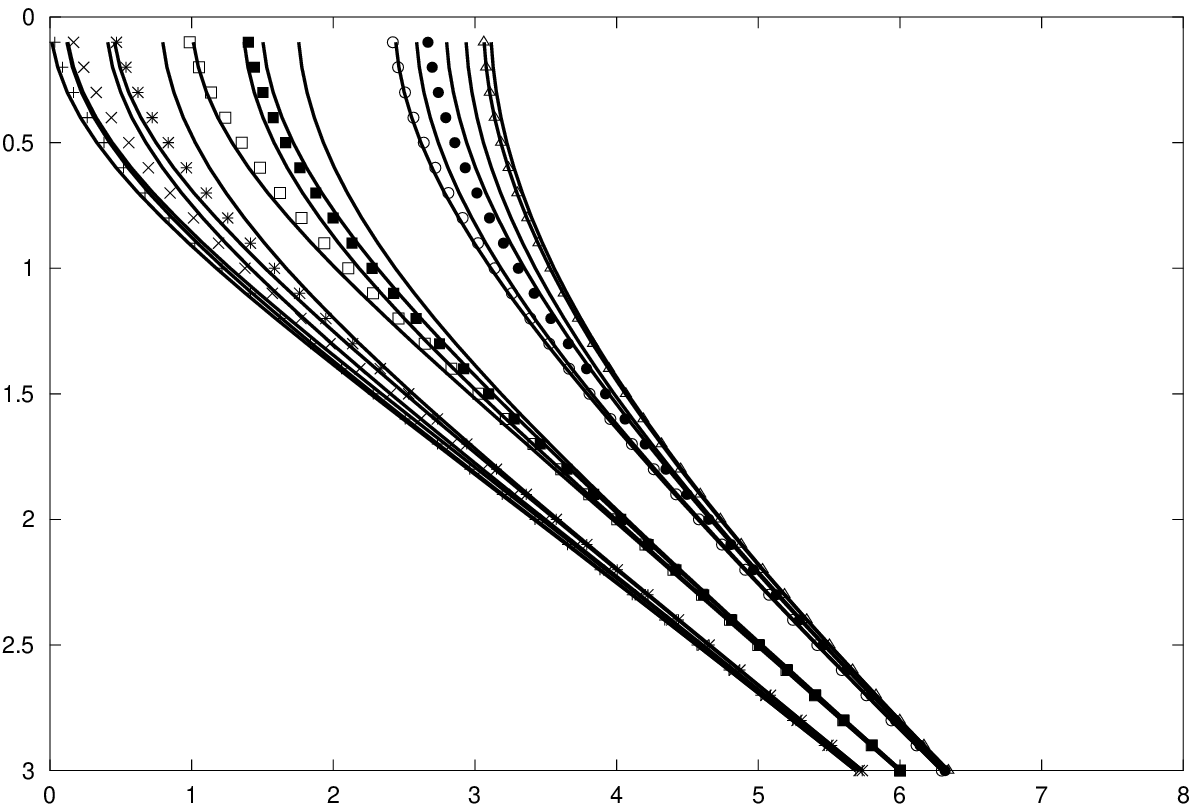}%
\end{picture}%
\setlength{\unitlength}{1973sp}%
\begingroup\makeatletter\ifx\SetFigFont\undefined
\def\x#1#2#3#4#5#6#7\relax{\def\x{#1#2#3#4#5#6}}%
\expandafter\x\fmtname xxxxxx\relax \def\y{splain}%
\ifx\x\y   
\gdef\SetFigFont#1#2#3{%
  \ifnum #1<17\tiny\else \ifnum #1<20\small\else
  \ifnum #1<24\normalsize\else \ifnum #1<29\large\else
  \ifnum #1<34\Large\else \ifnum #1<41\LARGE\else
     \huge\fi\fi\fi\fi\fi\fi
  \csname #3\endcsname}%
\else
\gdef\SetFigFont#1#2#3{\begingroup
  \count@#1\relax \ifnum 25<\count@\count@25\fi
  \def\x{\endgroup\@setsize\SetFigFont{#2pt}}%
  \expandafter\x
    \csname \romannumeral\the\count@ pt\expandafter\endcsname
    \csname @\romannumeral\the\count@ pt\endcsname
  \csname #3\endcsname}%
\fi
\fi\endgroup
\begin{picture}(11357,7615)(749,-9050)
\end{picture}

\end{center}

Figs 8-11: Band germs and bound or decaying states as functions
of $\gamma$ horizontal and $\beta$ vertical for Fibonacci strings
of 3-6 atoms. The bound states are not always inside the band germs.
The band germs which were glued in the periodic string are now
all separated.The number of band germs increases 
proportional to the number of atoms.
\vspace{0.2cm}

In Figs. 8-11 we use the roots of eqs. \ref{gl32}, \ref{gl33} to plot for the first 
$6$ Fibonacci strings  the edges of band germs 
and the bound states as functions of $\beta$, negative vertical 
coordinate, and $\gamma$, horizontal coordinate . For each value of
$m$, we compare the periodic case $q=1$ with the quasiperiodic case 
$q=\tau$. 

For the value $q=1$ we are back at periodic cases and would obtain glued 
systems of $2,3,5,8$ superband germs respectively in a way similar
to Figs. 5,6. These superband germs
encapsulate the bound states of the free clusters. For the 
Fibonacci case $q=\tau$ there appears a system of $f_m$ band germs, 
for any value of $\gamma$ separated by gaps. The bound states for
higher values of $\gamma$ are encapsulated within these band germs. 
For large fixed values of $\gamma$ the band germs become narrow and 
form a triple of  subsets. 
The latter feature may be related to the occurrence of isolated 
$2$-atom clusters at the short distance $b$ in the Fibonacci strings.
These may be considered as examples of motive clusters discussed in 
section 1.
The locally symmetric or antisymmetric hybrid states of these clusters
can be related to the lowest and highest subset.

Figs. 8-11 also demonstrate how the quasiperiodic system develops under
the Fibonacci recursion or inflation. If one looks for fixed strength
$\gamma$ at the interchange of bands and gaps, one observes a
permanent change of the density of states (DOS). In an approximant 
computation of this quasicrystal one would stop at a certain
length and obtain the DOS from the band germ analysis of the
corresponding string. In view of the present examples 
it is hard to believe that this method describes the real DOS of the
quasicrystal.\vspace{0.2cm}

In a periodic potential we get a finite number of bands at least at
negative energy which can encompass a number of electron states
proportional to the number of atoms.
In the Fibonacci strings notice that, in marked contrast to the periodic case,
when extending the Fibonacci string
the number of band germs which encapsulate bound states 
{\em increases proportional to the number of delta-potentials},
i.e. {\em proportional to the number} $n$ {\em of atoms}, each with one electron,
in the model. If this is
so, then the DOS cannot be computed from the gap and band germ 
structure of the string. As we have only $n$ electrons, it makes no sense
to distribute them into a system of $n$ bands. We conclude that 
in approximant computations of the DOS 
one should keep track of the number of bands in relation to the system size.
Otherwise the conclusions on the DOS in the proper quasiperiodic system are 
doubtful. \vspace{0.2cm}

Now we turn to the significance of the conserved quantity related to the
trace of the commutator. This trace is related to the invariant $I$ 
considered in \cite{KOH} and \cite{BAA} by
\begin{equation}
\label{gl33a}
\frac{1}{2} tr({\cal K}) = 2I+1.
\end{equation}
If the commutator becomes the $2 \times 2$ unit matrix we get $I=0$.
In \cite{KR3} we computed the commutator for the present
Fibonacci system with the result
\begin{equation}
\label{gl33b}
\frac{1}{2} tr({\cal K}) = 1+\frac{1}{2} (\frac{u}{\kappa})^2
(\sinh(\kappa(\tau-1)b)^2
\end{equation}
This expression is always larger than $1$. It implies that
{\em in the negative energy tight binding scenario the two transfer matrices never
commute}. A different result will appear for positive energy and scattering.

\section{Periodic strings at positive energy.}

We turn to electron states at positive energy. To this end we 
use the analytic continuation from the variable $\kappa$ to
the variable $k$ defined by
\begin{equation}
\label{gl34}
\kappa \rightarrow -ik.
\end{equation}
For the energy we have now $E=\frac{\hbar^2}{2m} k^2\geq 0$.
For the variables $\beta,\delta$ we shall use 
\begin{equation}
\label{gl35}
\beta \rightarrow -i\beta,\; \delta \rightarrow i\delta.
\end{equation}
So we maintain the symbol $\beta$ as a real variable in what follows. 
We shall plot 
the positive energy range $E= \frac{\hbar^2}{2mb^2}\beta^2$
at negative values of the new real parameter $\beta$.
Inserting the analytic continuation into the transfer matrix 
$\tilde{M}_1$ eq. \ref{gl3} 
we find
\begin{eqnarray}
\label{gl36}
x_1 & = & \cos(\beta)-\frac{1}{2}\delta \sin(\beta), 
\\ \nonumber
y_1 & = & i\sin(\beta)+i\frac{1}{2}\delta \cos(\beta), 
\\ \nonumber
\lambda_1 & = & \exp(-i\beta)= \exp(-ikb).
\end{eqnarray}
Similar expressions apply for $\tilde{M}_2$.

\subsection{The ${\cal S}$-matrix.}

The transfer matrix of a finite string can be rationally related to
the scattering matrix ${\cal S}$. For this purpose we shall use the
transfer matrix in a form similar to eq. \ref{gl2} which 
corresponds to exponential functions. The analytic continuation eqs.
\ref{gl34},\ref{gl35} is made
both in $M(x)$ and with the matrix $R:= \frac{1}{\sqrt{i}} \tilde{R}$ 
used instead of $\tilde{R}$. We redefine for positive energy  
\begin{equation}
\label{gl36a}
\tilde{M}(x):= R^{-1} M(x) R.  
\end{equation}
The negative energy tunnels with $\tilde{M}$ given by eq. \ref{gl2a}
become positive energy channels. The new fundamental system
of solutions is 
\begin{equation}
\label{gl36b}
M(x) R = \sqrt{\frac{1}{2k}} \left[
\begin{array}{ll}
\exp(ik x)& \exp(-ik x) \\
ik \exp(ik x)& -ik \exp(-ik x) \\
\end{array} \right].
\end{equation}
and consists of free plane waves running to the left and right
respectively. The transfer matrix in the channels becomes
\begin{eqnarray}
\label{gl36c}
\tilde{M}(x) &:= & R^{-1}M(x)R \\ \nonumber
             & = & \left[ 
\begin{array}{ll}
\exp (ik x) & 0 \\
0 & \exp (-ik x) \\
\end{array} \right].
\end{eqnarray}
Consider now a finite string of length $h$ with transfer matrix $\tilde{M}$
and with free channels on the left and right respectively.
Denote the amplitudes on the left and right of the string
by $l_+,l_-$ and $r_+,r_-$ respectively. These amplitudes are 
related by
\begin{equation}
\label{gl36d}
\left[
\begin{array}{l}
r_+ \\
r_- \\
\end{array}
\right]
= \left[
\begin{array}{ll}
a & b \\
c & d \\
\end{array}
\right]
\left[
\begin{array}{l}
l_+ \\
l_- \\
\end{array}
\right].
\end{equation}
The elements of the scattering matrix are now determined by 
the ratio of amplitudes under certain boundary conditions
and by phase factors which account for the length $h$ of the
string.
For {\em scattering from the left} we obtain
\begin{eqnarray}
\label{gl36e}
r_- & = &cl_++dl_-  =  0,
\\ \nonumber
S_{++} & = & \frac{r_+}{l_+} \exp(-ikh)
\\ \nonumber
& = & d^{-1} \exp(-ikh),
\\ \nonumber 
S_{-+} & = & \frac{l_-}{l_+}
\\ \nonumber
& = & - d^{-1} c.
\\ \nonumber 
\end{eqnarray}

For {\em scattering from the right} we get
\begin{eqnarray}
\label{gl36f}
l_+ & = & dr_+-br_-  =  0,
\\ \nonumber
S_{--} & = & \frac{l_-}{r_-} \exp(-ikh)
\\ \nonumber
& = & -d^{-1} \exp(-ikh),
\\ \nonumber 
S_{+-} & = & \frac{r_+}{r_-} \exp(-2ikh)
\\ \nonumber
& = & - d^{-1} b \exp(-2ikh).
\\ \nonumber 
\end{eqnarray}
Here $S_{++}$ and $S_{-+}$ are  the amplitudes of {\em forward} and of
{\em backward scattering} respectively. 
These expressions are given and discussed in \cite{KR1} eqs.(50-54).

${\bf 12\, Prop}$: The scattering matrix ${\cal S}$ has elements
which are rational expressions in the transfer matrix $\tilde{M}$.
The full scattering matrix  becomes
\begin{eqnarray}
\label{gl37}
{\cal S} & := &
\left[
\begin{array}{ll}
S_{++} & S_{+-} \\
S_{-+} & S_{--} 
\end{array}
\right]
\\ \nonumber
& = & 
\left[
\begin{array}{ll}
d^{-1}\exp(-ikh) & d^{-1}b\exp(-2ikh) \\
-d^{-1}c & d^{-1}\exp(-ikh) 
\end{array}
\right].
\end{eqnarray}
The exponential factors arise by requiring that for the free
transfer matrix  on the string of length $h$ we must have 
${\cal S}=1$. The expression for the ${\cal S}$-matrix in terms of the transfer
matrix is non-linear but rational. We must compute ${\cal S}$ from the transfer
matrix of the full string, or otherwise evaluate all the backscattering 
effects from the  components of the string.
The unitarity of the ${\cal S}$-matrix and its transformation under time
reversal follow from properties of 
the transfer matrix \cite{KR1}. 

Consider now the analytic 
continuation eq. \ref{gl34} in the inverse form 
\begin{equation}
\label{gl38}
k \rightarrow i\kappa,\; \kappa \geq 0.
\end{equation}
From inspection of eq. \ref{gl37} under this analytic continuation 
we find

${\bf 13\, Prop}$: The poles of the scattering matrix eq. \ref{gl37} on the positive
imaginary $k$-axis are given by $d(\kappa)=0$. These poles
determine the bound states 
of the system, fully in line with the analysis given in eq. \ref{gl33}.

\subsection{The ${\cal S}$-matrix for the periodic string $S^n$.}

Again we consider a range of positive energy such that $|x_1| \leq 1$.
This range may be called a band germ although it cannot correspond to bound
states of the system. In this range we can find the eigenvalues and 
eigenstates of $\tilde{M}_1$ by analytic continuation of the expressions
eqs. \ref{gl4}, \ref{gl5} . For the periodic string $S^n$ with the transfer matrix 
$\tilde{M}_1^n$ we find the same expressions as in eq. \ref{gl18}, but
with $y_1$ and $\lambda_1$ now taken from $\tilde{M}_1$ in eq. \ref{gl36}.
We shall again introduce a band label $K$ by 
\begin{equation}
\label{gl40}
\cos(Kb)= x_1 = \cos(\beta)-\frac{1}{2}\delta \sin(\beta).
\end{equation}
Consider first the backward scattering determined from eq. \ref{gl37}
by $c_n$ and $d_n$. The matrix element $c_n$ oscillates rapidly with large 
$n$. We obtain {\em maximum backscattering}  at the discrete values of the $K$-label
\begin{eqnarray}
\label{gl41}
Kb & = & \mu \pi, \mu = 0, \pm 1,
\\ \nonumber
c_n  & = & -i\frac{1}{2} \lambda_1^{-1} (-1)^{\mu (n-1)}n\delta, 
\\ \nonumber
d_n  & = & (-1)^{\mu n}(1-ny_1(-1)^{\mu}),
\\ \nonumber
\exp(-ikh) & = & \exp(-inkb).
\end{eqnarray}
From 
\begin{equation}
\label{gl42}
\cos(\mu \pi) = (-1)^{\mu},
\end{equation}
these maxima  correspond to the {\em edges of the positive energy band germs}
and become sharper with increasing $n$.

A {\em high-energy limit} is obtained for $\beta = kb \gg 1$. With
$\delta=\gamma/\beta$, eq. \ref{gl40} enforces 
\begin{equation}
\label{gl43}
\beta = kb = Kb +\nu 2\pi+\epsilon.
\end{equation}
Together with eq. \ref{gl41}, this equation yields {\em discrete values 
of $k$ on the points of the 
reciprocal lattice}. 

${\bf 14\,Prop}$: For large $n$ and sufficiently  high energy, the
backscattering amplitude takes its non-vanishing 
values on the points of the reciprocal
lattice.

To lowest order in $\epsilon$ we find 
\begin{eqnarray}
\label{gl44a}
x_1 & = & (-1)^{\mu} , \label{gl44} \\ \nonumber
y_1 & = & i(-1)^{\mu}\frac{1}{2}\delta,\\ \nonumber
d_n   & = & (-1)^{\mu n}(1-i\frac{1}{2}n\delta).
\end{eqnarray}
In the high-energy limit we then get from eq. \ref{gl37} at the band edges
\begin{eqnarray}
\label{gl45}
S_{++} & = & 
\frac{1}{1-i\frac{1}{2}n\delta}, \\ \nonumber
S_{-+} & = &  
\frac{i\frac{1}{2}n\delta}{1-i\frac{1}{2}n\delta}.
\label{gl45a}
\end{eqnarray}

The limits $n \rightarrow \infty$ and of large energy should be 
distinguished from one another. In the limit $n \rightarrow \infty$
the absolute value of the backward scattering amplitude approaches 
its maximum $|S_{-+}| =1$ allowed by unitarity, while the forward scattering 
amplitude goes to zero.

\section{Quasiperiodic strings at positive energy.}

The analytic continuation of the transfer matrix works 
for the Fibonacci strings. We simply must start the string with 
the analytic continuation of the starting matrices 
$\tilde{M}_1,\tilde{M}_2$ as given for $\tilde{M}_1$ in eq.\ref{gl36}.
Then we can apply the recursion technique of eq. \ref{gl30}
to find an algebraic expression for the transfer matrix
of the Fibonacci string $W_m$.

\subsection{The Fibonacci-atlas.}

In principle we can construct the elements of the transfer matrix as
polynomials of degree $f_m$ with respect to $\delta$ for any order $m$. 
We  obtain the corresponding ${\cal S}$-matrix from the relations 
eq. \ref{gl37}.
In general it will be hard to obtain in this way  closed limiting expressions
as was possible in the periodic case. Nevertheless there are particular
regions where closed expressions can be derived. We illustrate these 
regions in Fig.12 by plotting the band germs of both transfer matrices
$\tilde{M}_1,\tilde{M}_2$ for $q=\tau$ 
as functions of the variables $\beta,\gamma$
in the range $\beta \geq 0, \gamma >0,\gamma<0$. This plot we call 
the {\em Fibonacci-atlas}.

\begin{center}
\begin{picture}(0,0)%
\includegraphics{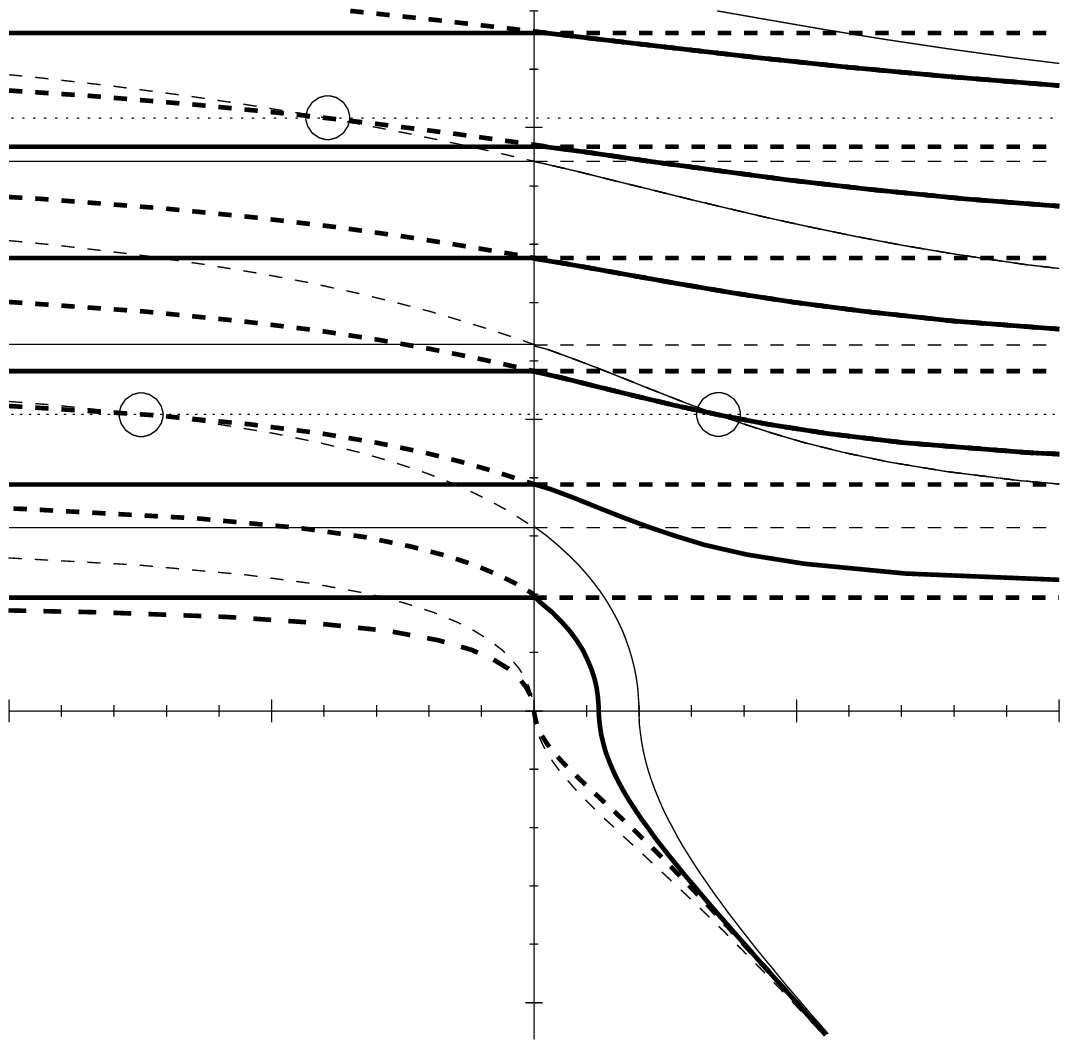}%
\end{picture}%
\setlength{\unitlength}{2763sp}%
\begingroup\makeatletter\ifx\SetFigFont\undefined
\def\x#1#2#3#4#5#6#7\relax{\def\x{#1#2#3#4#5#6}}%
\expandafter\x\fmtname xxxxxx\relax \def\y{splain}%
\ifx\x\y   
\gdef\SetFigFont#1#2#3{%
  \ifnum #1<17\tiny\else \ifnum #1<20\small\else
  \ifnum #1<24\normalsize\else \ifnum #1<29\large\else
  \ifnum #1<34\Large\else \ifnum #1<41\LARGE\else
     \huge\fi\fi\fi\fi\fi\fi
  \csname #3\endcsname}%
\else
\gdef\SetFigFont#1#2#3{\begingroup
  \count@#1\relax \ifnum 25<\count@\count@25\fi
  \def\x{\endgroup\@setsize\SetFigFont{#2pt}}%
  \expandafter\x
    \csname \romannumeral\the\count@ pt\expandafter\endcsname
    \csname @\romannumeral\the\count@ pt\endcsname
  \csname #3\endcsname}%
\fi
\fi\endgroup
\begin{picture}(7362,7098)(772,-8023)
\put(4266,-7821){\makebox(0,0)[lb]{\smash{\SetFigFont{8}{9.6}{bf}$5$}}}
\put(6217,-6036){\makebox(0,0)[lb]{\smash{\SetFigFont{8}{9.6}{bf}$10$}}}
\put(7992,-6036){\makebox(0,0)[lb]{\smash{\SetFigFont{8}{9.6}{bf}$20$}}}
\put(772,-6036){\makebox(0,0)[lb]{\smash{\SetFigFont{8}{9.6}{bf}$-20$}}}
\put(2527,-6036){\makebox(0,0)[lb]{\smash{\SetFigFont{8}{9.6}{bf}$-10$}}}
\put(4605,-1836){\makebox(0,0)[lb]{\smash{\SetFigFont{8}{9.6}{bf}$-10$}}}
\put(4595,-3871){\makebox(0,0)[lb]{\smash{\SetFigFont{8}{9.6}{bf}$-5$}}}
\end{picture}

\end{center}

Fig.12: Fibonacci atlas: Band germs of the transfer matrices 
$\tilde{M}_1,\tilde{M}_2$
for positive and negative energy as functions of the energy variable 
$\beta$ and the strength variable $\gamma$.
Upper band edges drawn as full lines, lower band edges as dashed lines,
heavy lines for the matrix $\tilde{M}_2$.
The two weak horizontal dotted lines mark values $\beta(p)$ where the two transfer matrices
commute and give rise to extended states.\vspace{0.2cm}

Consider again the commutator ${\cal K}$ of the two transfer matrices.
We must insert the analytic continuation to positive energy and obtain from
eq. \ref{gl33b}
\begin{equation}
\label{gl46}
\frac{1}{2} tr(K) = 1+\frac{1}{2} (\frac{u}{k})^2
(\sin(k(\tau-1)b)^2
\end{equation}
This expressions shows that there are periodic points where the
half-trace becomes $1$ and the invariant becomes $I=0$. These
points are given by
\begin{equation}
\label{gl47}
\beta(p)  = \tau p \pi,\; p= 0,1,\ldots.
\end{equation}
Moreover it was shown in \cite{BAA} that the deviations 
$I-1$ 
at these points starts quadratically with $(\beta-\beta(p))$.
The values $\beta(p)$ are marked by dotted horizontal lines in Fig.12. 
If a line of this type intersects, depending on $\gamma$, with
overlapping band germs for both transfer matrices, 
we obtain 
from $\lambda_2 = (-1)^p \lambda_1$ their 
proportionality, 
\begin{equation}
\label{gl48}
\tilde{M}_2 = (-1)^p \tilde{M}_1. 
\end{equation}
${\bf 15\, Prop}$: At the values eq. \ref{gl46}, the two transfer 
matrices $\tilde{M}_1,\tilde{M}_2$ commute and are even proportional
to one another.

More precisely we require three conditions at these special points:
The half-traces of both transfer matrices should be smaller than $1$,
and the commutator should become the unit matrix.
These conditions are controlled by the Fibonacci atlas Fig. 12:
Consider a dotted horizontal line corresponding to commuting transfer 
matrices.  For not too large  values of $\gamma$, 
it runs inside overlaps of the two band germs which
imply that both $|x_1| \leq 1,|x_2| \leq 1$. Outside this region,
that is outside the points 
marked by circles in Fig. 12, we still have commuting transfer 
matrices, but their individual half-traces are larger than $1$.
It follows from algebraic properties of the group $SU(1,1)$ given in 
\cite{KR1} that a vanishing  commutator implies group elements
of the same class type.

We can now see the advantage of the present algebraic and polynomial 
approach 
compared to numerical studies of the trace systems:
For example in  \cite{BAA} the points eq. \ref{gl47} were explored  
numerically and  for fixed
values of a strength $\gamma<0 $,  corresponding to repulsive 
delta-potentials. The Fibonacci-atlas of Fig.12 now 
yields from closed algebraic expressions  a full view on the 
regions of commutativity for all positive
and negative values of $\gamma$ and positive energy.

In a neighbourhood of such a commutative point we get, 
using the commutativity,
for the full Fibonacci string
$W_m$ the transfer matrix 
\begin{equation}
\label{gl49}
\tilde{M}_m = (-1)^{pf_{m-1}} \tilde{M}_1^{f_m}.
\end{equation}
This transfer matrix agrees up to a factor with the transfer matrix of
a periodic string of $f_m$ delta-potentials with the spacing $b$.
It follows that the ${\cal S}$-matrix of the Fibonacci strings 
near these points has the form and limiting values discussed in
section 4.2. 

${\bf 16\,Prop}$: In the neighbourhood of the discrete positive energies 
corresponding to eq. \ref{gl47}, the transfer and ${\cal S}$-matrix 
of the Fibonacci system are equivalent to those of a periodic 
string of length $f_mb$. 

Some care is required because
$\beta(p)$ in eq. \ref{gl47} is not rational whereas for the periodic
case we used the discrete labels $Kb$ and $\beta=kb$.
Note however that the value $\beta(p)$ for $p=f_l$ 
and not too small values of
$l$ is well approximated by the integral multiple 
$\beta = f_{l+1}\pi$ of $\pi$. 

Finally in Fig. 13 we give the real wave function of an electron
travelling at positive energy through a Fibonacci string of
attractive delta potentials. The energy is tuned to a commutative
value. At each passage through a delta
potential, the derivative of the wave function jumps by a finite
value.

\begin{center}
\begin{picture}(0,0)%
\includegraphics{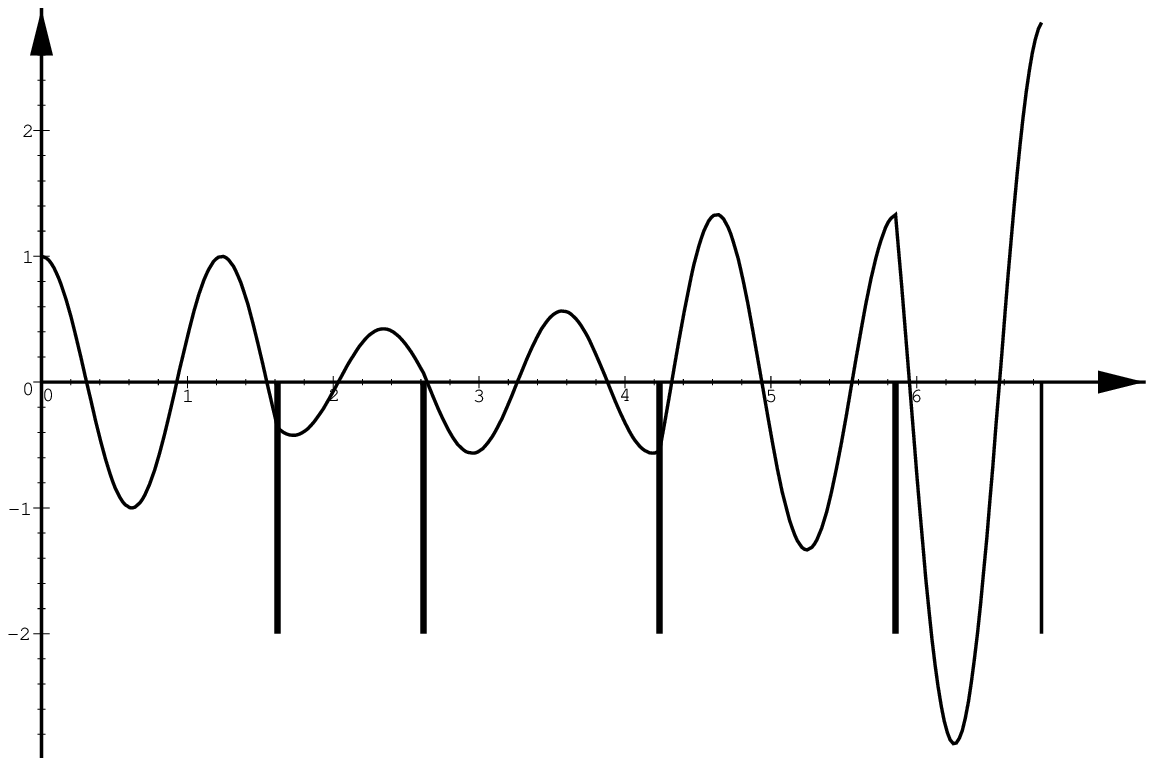}%
\end{picture}%
\setlength{\unitlength}{1973sp}%
\begingroup\makeatletter\ifx\SetFigFont\undefined
\def\x#1#2#3#4#5#6#7\relax{\def\x{#1#2#3#4#5#6}}%
\expandafter\x\fmtname xxxxxx\relax \def\y{splain}%
\ifx\x\y   
\gdef\SetFigFont#1#2#3{%
  \ifnum #1<17\tiny\else \ifnum #1<20\small\else
  \ifnum #1<24\normalsize\else \ifnum #1<29\large\else
  \ifnum #1<34\Large\else \ifnum #1<41\LARGE\else
     \huge\fi\fi\fi\fi\fi\fi
  \csname #3\endcsname}%
\else
\gdef\SetFigFont#1#2#3{\begingroup
  \count@#1\relax \ifnum 25<\count@\count@25\fi
  \def\x{\endgroup\@setsize\SetFigFont{#2pt}}%
  \expandafter\x
    \csname \romannumeral\the\count@ pt\expandafter\endcsname
    \csname @\romannumeral\the\count@ pt\endcsname
  \csname #3\endcsname}%
\fi
\fi\endgroup
\begin{picture}(11905,7264)(601,-7744)
\put(12001,-4711){\makebox(0,0)[lb]{\smash{\SetFigFont{10}{12.0}{bf}$x$}}}
\put(601,-1186){\makebox(0,0)[lb]{\smash{\SetFigFont{10}{12.0}{bf}$\psi(x)$}}}
\end{picture}

\end{center}

Fig. 13: Electron state propagating through a Fibonacci string
of attractive delta-potentials at a positive energy
where the transfer matrices commute.\vspace{0.2cm}

\section{Conclusion.}

Our study of a very simple $1D$ electron system has demonstrated 
most of the points stated in general terms in the introduction,
section 1. The local boundary conditions were implemented and
related. The negative and positive energy scenarios showed
different behaviour. At negative energy, the bound states 
have a clear relation to the Bloch states. In going from
periodic to quasiperiodic strings, the glued and gap-less systems
of superbands open gaps and split into subbands which encapsulate bound 
states. Motive clusters appear in the bound state energy.
With increasing length of the Fibonacci string, 
the subband structure changes 
in a non-trivial fashion, with a number of subbands proportional
to the number of atoms,  and so puts question marks to approximant
calculations. At positive energy, the band edges in periodic strings
are related to maxima of  scattering amplitudes. For quasiperiodic 
Fibonacci strings we gave closed algebraic expressions. At 
positive energies related by a period wrt. $k$, the transfer matrices
commute and give rise to maxima in the scattering amplitudes.

\end{document}